\documentclass[aps,prb, twocolumn, preprintnumbers,amsmath,amssymb,superscriptaddress, floatfix]{revtex4}
\usepackage{simplewick}
\usepackage{graphicx}
\usepackage{latexsym}
\usepackage{amsmath}
\usepackage{graphics}
\usepackage{amssymb}
\usepackage{layout}
\usepackage{verbatim}
\usepackage{amsfonts,epsfig}
\newcommand{\ket}[1]{\left|#1\right>}
\newcommand{\bra}[1]{\left< #1 \right|}
\newcommand{\beq}{\begin{equation}}
\newcommand{\eeq}{\end{equation}}
\newcommand{\HH}{\hat{H}}
\newcommand{\mean}[1]{\langle{#1}\rangle{}}
\newcommand{\Vsf}{\hat{V}_{\text{sf}}}
                                                                                                                                                                                                                                       
\begin{document}
\title{Pure quantum dephasing of a solid state electron spin qubit in a large nuclear spin bath
coupled by long-range hyperfine-mediated interactions}
\author{{\L}ukasz Cywi{\'n}ski}
\affiliation{Condensed Matter Theory Center, Department of Physics, University of Maryland, College Park, MD 20742-4111, USA}
\affiliation{Institute of Physics, Polish Academy of Sciences, al.~Lotnik{\'o}w 32/46, PL 02-668 Warszawa, Poland}
\author{Wayne M. Witzel}
\affiliation{Condensed Matter Theory Center, Department of Physics, University of Maryland, College Park, MD 20742-4111, USA}
\affiliation{Naval Research Laboratory, Washington, DC 20375, USA}
\affiliation{Sandia National Laboratories, Albuquerque, New Mexico 87185, USA}
\author{S. Das Sarma}
\affiliation{Condensed Matter Theory Center, Department of Physics, University of Maryland, College Park, MD 20742-4111, USA}
\date{\today }

\begin{abstract}
We investigate decoherence due to pure dephasing of a localized spin qubit interacting with a nuclear spin bath. Although in the limit of a very large magnetic field the only decoherence mechanism is spectral diffusion due to dipolar flip-flops of nuclear spins, with decreasing field the hyperfine-mediated interactions between the nuclear spins become important. We take advantage of their long-range nature,  and resum the leading terms in an $1/N$ expansion of the decoherence time-evolution function ($N$, being the number of nuclear spins interacting appreciably with the electron spin, is large). For the case of the thermal uncorrelated bath we show that our theory is applicable down to low magnetic fields ($\sim\!10$ mT for a large dot with $N\! = \! 10^{6}$) allowing for comparison with recent experiments in GaAs quantum dot spin qubits.
Within this approach we calculate the free induction decay and spin echo decoherence in GaAs and InGaAs as a function of the number of the nuclei in the bath (i.e.~the quantum dot size) and the magnetic field. Our theory for free induction decay in a narrowed nuclear bath is shown to agree with the exact solution for decoherence due to hyperfine-mediated interaction which can be obtained when all the nuclei-electron coupling constants are identical. For the spin echo evolution we show that the dominant decoherence process at low fields is due to interactions between nuclei having significantly different Zeeman energies (i.e.~nuclei of As and two isotopes of Ga in GaAs), and we compare our results with recent measurements of spin echo signal of a single spin confined in a GaAs quantum dot.  
For the same set of parameters we perform calculations of decoherence under various dynamical decoupling pulse sequences, and predict the effect of these sequences in low $B$ regime in GaAs.
\end{abstract}

\maketitle

\section{Introduction}
The quantum dynamics of a localized central spin coupled to an environment of other fluctuating spins (spin-bath) is probably the longest studied model of quantum decoherence, outgrowing from research on dephasing and relaxation of ensembles of spins in Nuclear Magnetic Resonance.\cite{Slichter,Abragam} 
In recent years this field of research has been reinvigorated by experimental progress in creation, initialization, readout, and coherent control of electron spin qubits in semiconductors.\cite{Hanson_RMP07,Cerletti_Nanotech05,Elzerman_Nature04,Braun_PRL05,Dutt_PRL05,Koppens_Science05,Johnson_Nature05,Petta_Science05,Atature_Science06,Koppens_PRL08,Childress_Science06,Hanson_Science08,Greilich_NP09} 
Understanding the decoherence caused by the nuclear spin-bath is crucial for development of  these qubits, since at low temperatures and moderate magnetic fields all the other sources of electron spin relaxation and dephasing (e.g.~spin-orbit scattering with phonons) are strongly suppressed.\cite{Khaetskii_PRB01,Erlingsson_PRB02,Golovach_PRL04,Semenov_PRL04}
The remaining relevant interaction is the hyperfine (hf) interaction of an electron spin with the nuclear spins of the host material. In quantum computing architectures based on III-V materials such as GaAs and InAs the presence of the nuclear bath is unavoidable, since all the isotopes of Ga, As, and In have non-zero nuclear spins.\cite{Schliemann_JPC03} The problem at hand is hard, since the bath is interacting: the nuclei are effectively interacting due to their mutual hf coupling to the ``central spin'' of the electron, and furthermore they are also coupled by ``intrinsic'' intra-bath dipolar interactions.\cite{Schliemann_JPC03,deSousa_PRB03} 
The resulting dynamics of the electron is of highly nontrivial non-Markovian character,\cite{Coish_PRB04} since time-scales of the bath dynamics are orders of magnitude longer than the typical electron spin precession and decoherence time-scales. A simple perturbation theory in hf interaction quickly breaks down,\cite{Khaetskii_PRL02,Khaetskii_PRB03} and one has to use more global approaches (e.g.~involving infinite-order resummations of the perturbation theory) in order to obtain physical results.
However, significant theoretical progress has been made with a combination of analytical work \cite{Khaetskii_PRL02,Khaetskii_PRB03,deSousa_PRB03,Coish_PRB04,Yao_PRB06,Liu_NJP07,Witzel_PRB05,Witzel_PRB06,Witzel_PRB08,Saikin_PRB07,Deng_PRB06,Deng_PRB08,Coish_PRB08,Yang_CCE_PRB08,Cywinski_PRL09,Yang_CCE_PRB09} and exact numerics for small systems.\cite{Schliemann_JPC03,Dobrovitski_PRE03,Shenvi_scaling_PRB05,Zhang_PRB06,Zhang_JPC07,Zhang_Dobrovitski_PRB07,Zhang_Viola_PRB08} 
A successful comparison between the theory\cite{deSousa_PRB03,Witzel_PRB05,Witzel_PRB06,Witzel_AHF_PRB07,Saikin_PRB07} and experiment\cite{Tyryshkin_PRB03,Tyryshkin_JPC06} has been made for Si:P spin qubits (based on the spin of electron bound to the phosphorus donor). These calculations have shown that in experiments on Si:P system the decoherence is due to the dipolar interactions between the nuclei - the so-called spectral diffusion (SD) process, in which the hf interaction serves only as a transmitter of the dipolar-induced nuclear dynamics to the electron. The analogous calculations for GaAs dots were only performed\cite{Witzel_PRB06,Yao_PRB06,Witzel_PRB08} for magnetic fields much higher than $B\approx 0.1$ T used in experiments on decoherence of a single spin in a GaAs dot,\cite{Koppens_PRL08} and they gave time-scales two orders of magnitude longer than the observed\cite{Koppens_PRL08} decay in $\approx 0.3\, \mu$s. 
This clearly showed that decoherence at low field in GaAs is \emph{not} due to spectral diffusion.
At fields currently experimentally relevant for III-V based spin qubits one clearly has to consider the nuclear dynamics caused by the hf interaction, specifically by the nuclei-nuclei interactions due to virtual electron-nuclear spin flips.\cite{Yao_PRB06,Liu_NJP07,Saikin_PRB07,Coish_PRB08} In this paper we present the detailed derivation of the theory\cite{Cywinski_PRL09} of decoherence due to these hf-mediated interactions which is applicable down to $\sim \! 10$ $(100)$ mT fields in quantum dots containing $\sim \! 10^{6}$ $(10^{4})$ nuclei.

The hf interaction is given by
\beq
\hat{H}_{\text{hf}} = \sum_{i} A_{i} \mathbf{S}\cdot \mathbf{J}_{i} = \sum_{i} A_{i} [ \hat{S}^{z}\hat{J}^{z}_{i} + \frac{1}{2} ( \hat{S}^{+}\hat{J}^{-}_{i} + \hat{S}^{-}\hat{J}^{+}_{i} )  ] \,\, ,  \label{eq:hf}
\eeq
where $A_{i}$ is the hf coupling of the $i$-th nuclear spin to the electron spin.
This interaction causes both the energy relaxation of the qubit (decay of the spin component parallel to the magnetic field) and its dephasing (decay of components of the spin transverse to the magnetic field). The energy relaxation is strongly suppressed by a finite magnetic field:\cite{Khaetskii_PRL02,Khaetskii_PRB03,Coish_PRB04} due to the large mismatch of electronic and the nuclear Zeeman energies the hf interaction is not efficient at relaxing the $z$ component of the spin at fields larger than $\sim\! 10$ mT in GaAs quantum dots. More precisely, the longitudinal polarization loss was shown\cite{Coish_PRB04} to be of the order of $\delta^{2}$, with the small parameter $\delta$ defined as
\beq
\delta \equiv \frac{\mathcal{A}}{\Omega\sqrt{N}} \,\, ,  \label{eq:delta}
\eeq
where $\mathcal{A}=\sum_{i}A_{i}$ is the total hf interaction energy, $\Omega$ is the electron spin Zeeman splitting, and $N$ is the number of nuclei interacting appreciably with the electron. In quantum dots the value of $N$ ranges from $10^{4}$ in smallest self-assembled dots\cite{Dutt_PRL05,Greilich_Science06} to $10^{6}$ in lateral gated dots,\cite{Petta_Science05,Koppens_PRL08} and typical values of $\mathcal{A}$ and $\Omega$ are given in Table \ref{tab:energies}.

The dephasing of the qubit, on the other hand, while being detrimental for quantum computation applications, is much harder to suppress. 
Theoretically, the simplest possible way to eliminate dephasing is to spin-polarize the nuclear bath, but the $> \! 90$\% polarization necessary to achieve a meaningful increase in coherence time\cite{Coish_PRB04,Deng_PRB06,Zhang_PRB06} is impossible to achieve experimentally.\cite{Baugh_PRL07,Reilly_Science08,Petta_PRL08} Here we focus on  electron spin dephasing due to hf interaction with an unpolarized nuclear bath. Specifically, we consider a thermal uncorrelated state of the nuclei, which due to the smallness of the intrinsic nuclear energy scale (nuclear Zeeman splitting is $\sim$ mK at $B$ fields of the order of a Tesla) means that we will use the nuclear density matrix proportional to unity. Let us note that specially designed experimental protocols have been shown to drive the nuclear bath into a different state, characterized by a small net polarization and, more importantly, a ``narrowed'' distribution of  nuclear polarizations.\cite{Reilly_Science08} 

% FID T2
One operational definition of the spin dephasing time is the characteristic time of Free Induction Decay (FID) of the spin.\cite{Slichter,Abragam} In the FID experiment the spin is oriented in the $xy$ plane at $t=0$ with the field $\mathbf{B} \parallel \hat{z}$ giving the major part of the electron precession frequency, and the subsequent precession and decay of the average in-plane spin is measured. It is crucial to note that the electron precession frequency consists of not only the Zeeman frequency $\Omega$, but also the Overhauser field $\sum A_{i}J^{z}_{i}$ arising from all the nuclear spins.
The current FID experiments are either spatial\cite{Dutt_PRL05,Greilich_Science06} or temporal ensemble\cite{Johnson_Nature05,Petta_Science05}  measurements. In the first case one addresses many electron spins at the same time, each one of them interacting with a different nuclear bath. In the second case one measures repeatedly the same spin, but the total data acquisition time is long enough for the nuclear bath to appreciably change its state and probe its phase space. 
These two situations involving spatial and temporal averaging are presumably equivalent from a statistical perspective, leading to the same characteristic FID time.
In both cases, the experimental procedure translates into averaging over the relevant density matrix of the nuclei. The rms of the Overhauser field is $\sim \mathcal{A}/\sqrt{N}$ in a thermal bath, leading to a strong inhomogeneous broadening of the FID signal.\cite{Merkulov_PRB02} 
The resulting observed decay is Gaussian, i.e.~the in-plane spin decays as $\exp[-(t/T^{*}_{2})^2 ]$, with a characteristic time-constant of $T^{*}_{2} \sim \sqrt{N}/\mathcal{A}$ being the order of $10$ ns in III-V quantum dots.\cite{Dutt_PRL05,Johnson_Nature05,Petta_Science05} 

The $T^{*}_{2}$ decay of FID is a purely classical averaging phenomenon, which does not even require the bath to have any dynamics. The presence of strong inhomogeneous broadening underlines the necessity of control of the random Overhauser shifts of many qubits in a spin-based quantum computer architecture, but the actual coherence of a qubit is characterized by a ``single-spin'' FID decay time $T_{2}$. 
This is the decay time which one would measure for a single spin if the duration of the experiment  was shorter than the time on which the Overhauser field fluctuates appreciably, or for an ensemble of spins having the same Overhauser shift. In the latter case one would be dealing with a so-called narrowed state of the nuclei. 
There are a few theoretical proposals on how to drive the nuclei into such a narrowed state\cite{Klauser_PRB06,Giedke_PRA06,Stepanenko_PRL06} and experimental progress has been made.\cite{Greilich_Science06,Greilich_Science07,Reilly_Science08} 
In theoretical calculations of FID decoherence\cite{Coish_PRB04,Yao_PRB06,Liu_NJP07,Deng_PRB06,Deng_PRB08,Coish_PRB08} the maximally narrowed states are used (i.e.~eigenstates of the Overhauser operator), and it is argued that in large bath the decoherence dynamics does not depend on the choice of the state, with the tacit assumption that the used state is a ``typical'' one (i.e.~not one of the few highly polarized low-entropy states). In this paper we take an analogous approach, and below by FID decoherence we refer to the ``single-spin'' process.

% SPIN ECHO GENERAL
Whereas the ideas for state narrowing of nuclei are very recent, for the past 50 years the preferred way of removing the inhomogeneous broadening has been  the spin echo (SE) experiment,\cite{Slichter,Abragam} in which the electron spin is rotated by angle $\pi$ around one of the in-plane axes at the midpoint of its evolution. Such a protocol can be denoted as $t/2-\pi-t/2$: two free evolution periods with a fast external pulse in the middle and readout at final time $t$. It is easy to see that such a procedure will remove inhomogeneous broadening (static spread of the precession frequencies). In NMR literature the SE decay time is often identified with single-spin $T_{2}$, but this holds only when the decoherence due to the actual dynamics of the spin and the bath can be treated in a Markovian approximation (leading to $\exp(-t/T_{2})$ decay). This is generally not the case for the spin interacting with a nuclear bath due to the slowness of the bath dynamics and back-action of the electron spin on the bath,\cite{Coish_PRB04,Yao_PRB06} and the Markovian decay of spin coherence has been predicted\cite{Liu_NJP07,Coish_PRB08,Cywinski_PRL09} only in very specfic conditions (long times and large magnetic fields). Thus the characteristic decay time in the SE experiment, denoted by $T_{\text{SE}}$, is most often different from FID $T_{2}$ time. 
The SE is the simplest available method of uncovering the nontrivial quantum dynamics from underneath the inhomogeneous broadening, and it has been succesfully applied in Refs.~\onlinecite{Petta_Science05,Koppens_PRL08} to lateral GaAs quantum dot qubits and in Refs.~\onlinecite{Tyryshkin_PRB03,Tyryshkin_JPC06} to Si:P qubits (and very recently significant progress has been made towards repeating these experiments in InGaAs self-assembled dots,\cite{Greilich_NP09} where optical control of electron spin is employed). Consequently, a reliable method of calculating $T_{\text{SE}}$ is crucial for establishing contact between the experiments and theory. We note that the state narrowing protocols of Ref.~\onlinecite{Reilly_Science08} are unlikely to lead to a decay time shorter than that obtained through the spin echo technique.
 
 % SPIN ECHO DETAILS
For many years the theories of SE decay due to the spectral diffusion (i.e.~dipolar interactions) were based on stochastic assumptions (see Ref.~\onlinecite{deSousa_PRB03} and references therein) Only recently  have fully quantum mechanical and efficient approaches to the nuclear bath dynamics under the SE sequence been developed.\cite{Witzel_PRB05,Witzel_PRB06,Yao_PRB06,Liu_NJP07,Saikin_PRB07,Witzel_PRB08} All these theories involve some kind of exponential resummation of the perturbation series for the decoherence evolution function. The theoretical methods include expansion in real-space clusters,\cite{Witzel_PRB05,Witzel_PRB06,Witzel_PRB08} a related pair correlation approximation (PCA) approach,\cite{Yao_PRB06,Liu_NJP07,Yang_PRB08} and the diagrammatic linked-cluster technique.\cite{Saikin_PRB07} For an excellent discussion of the relation between these methods see Ref.~\onlinecite{Yang_CCE_PRB08}, where also a new ``cluster-correlation expansion'' is proposed for small nuclear baths, like the one interacting with the NV center in diamond.\cite{Hanson_Science08,Maze_PRB08}
These theories have shown a very good agreement\cite{Witzel_PRB06,Witzel_AHF_PRB07,Saikin_PRB07} with the spin echo experiments performed on Si:P system.\cite{Tyryshkin_PRB03,Tyryshkin_JPC06} However, these works have concentrated on a somewhat vaguely defined high-field regime, in which the SE decay is dominated by spectral diffusion, since the decoherence due to the hf-mediated flip-flop of pairs of nuclei is nearly completely removed by the SE sequence at $B \! \approx \! 10$ T.
This was first observed in numerical simulations,\cite{Shenvi_scaling_PRB05} and then explained in a physically transparent way within the PCA approach.\cite{Yao_PRB06,Liu_NJP07} However, this cannot explain the SE experiments\cite{Koppens_PRL08} in GaAs at $B \! \leq  \! 0.1$ T, underlining the necessity for a reliable low-field theoretical approach.

% DYNAMICAL DECOUPLING:
The calculation of the decoherence in the SE experiment is in principle qualitatively different than for FID, since one is considering the driven evolution of the system. Once the $\pi$ pulse driving is incorporated into the theoretical framework, one can generalize the calculation to the evolution driven by sequences of $\pi$ pulses more complicated than the SE.
The application of such \emph{dynamical decoupling} (DD) sequences is expected to prolong the coherence time.\cite{Viola_JMO04,Khodjasteh_PRA07,Uhrig_PRL07} The above-mentioned analytical theories were used to calculate this effect for the electron spin driven by  the classical Carr-Purcell-Meiboom-Gill sequence\cite{Witzel_PRL07}, by the concatenated DD,\cite{Yao_PRL07,Witzel_CDD_PRB07} and by Uhrig's DD sequence.\cite{Lee_PRL08} Numerical investigations of the very low-field regime (inaccessible to all the mentioned theories) have also been done.\cite{Zhang_Dobrovitski_PRB07,Zhang_Viola_PRB08}

% HISTORY OF THEORY focusing on hf interactions
Compared to the well-understood status of the spectral diffusion theory of decoherence due to dipolar interactions in the bath (relevant at large B), the relation between different analytical  theories of dephasing due to hf interactions alone is less-well established. In the first studies of the electron spin dynamics due to hf interaction, the processes of electron-nuclear spin flip were treated directly,\cite{Khaetskii_PRL02,Khaetskii_PRB03,Coish_PRB04} i.e.~the perturbation expansion was done in orders of interaction $\hat{V}_{\text{sf}} \sim ( \hat{S}^{+}\hat{J}^{-} + \hat{S}^{-}\hat{J}^{+})$, and in the second order of Generalized Master Equation (GME) approach no genuine dephasing (i.e.~decay to zero of the transverse spin) has been obtained.\cite{Coish_PRB04} The FID decay to zero was obtained from $\hat{V}_{\text{sf}}$ using the equations of motion method for spin correlation functions,\cite{Deng_PRB06,Deng_PRB08} and then by the GME approach in the fourth order with respect to $\hat{V}_{\text{sf}}$.\cite{Coish_PRB08} In the meantime, a physically intuitive PCA solution for both FID and SE decay has been proposed,\cite{Yao_PRB06,Liu_NJP07} and later derived using the diagrammatic linked cluster expansion.\cite{Saikin_PRB07,Yang_CCE_PRB08} In these papers the $\hat{V}_{\text{sf}}$ interaction was not treated directly, but an effective inter-nuclei interaction due to virtual electron-nuclear spin flips was used. 
This hf-mediated interaction approach was also used succesfully within the GME framework.\cite{Coish_PRB08} 
In all of the above papers, the condition of applicability of the theory was either explicitly or implicitly given as $\mathcal{A}/\Omega < 1$, which corresponds to $B$ larger than a few Tesla in GaAs. At the same time, it has been well established\cite{Coish_PRB04} that the real electron spin-flip is negligible as long as $\delta \! \ll \! 1$, which is a much less restrictive criterion. In fact, this criterion was suggested to be the relevant one in Ref.~\onlinecite{Yao_PRB06} where PCA method was introduced. Finally it should be mentioned that the limit of very low $B$ fields corresponding to $\delta \! > \! 1$ ($B$ less than a few mT in GaAs) is accessible only to exact numerical simluations.\cite{Zhang_PRB06,Zhang_JPC07,Zhang_Dobrovitski_PRB07,Zhang_Viola_PRB08}

% PURPOSE
In this paper we present a detailed derivation of a theory\cite{Cywinski_PRL09} of pure dephasing under any sequence of ideal $\pi$ pulses, which is based on the effective nuclear Hamiltonian approach in which the electron-nuclear spin flips are eliminated in favor of hf-mediated interactions between the nuclei.\cite{Shenvi_scaling_PRB05,Yao_PRB06,Coish_PRB08} 
We argue that for nuclear bath being unpolarized and uncorrelated (its density matrix being a thermal one) this theory is applicable at experimentally relevant low magnetic fields (when $\mathcal{A}/\Omega \! > \! 1$), with the controlling small parameter $\delta$ defined in Eq.~(\ref{eq:delta}). This result is firmly established only up to the time scale of $N/\mathcal{A} \sim 10 \, \mu$s (for $N \! = \! 10^{6}$), but at low $B$ most of the decay occurs on this time scale. The key insight leading to the solution of the decoherence problem is the long-range character of the hf-mediated interactions, which couple all the $N \! \sim \! 10^{5}-10^{6}$ nuclear spins appreciably interacting with the electron spin. This allows us to identify the so-called ring diagrams as the leading terms in the perturbation expansion of the decoherence function, with corrections to them suppressed as $1/N$. This is analogous to the summation of the ring diagrams in the cluster expansion of the partition function of the high-density Ising model well known from classical statistical mechanics.\cite{Brout_PR60} A similar approach (termed ``the pair correlator approximation'') has also been used recently to calculate the influence of the hf-mediated interaction on the electric dipole spin resonance in quantum dots.\cite{Rashba_PRB08}
Taking into account the ring diagrams only, we can sum all the terms in the cumulant (linked cluster) expansion of the decoherence function, obtaining closed formulas for decoherence in FID, SE, and other experiments involving more $\pi$ pulses. 
We compare our FID results with the experiments on InGaAs dots,\cite{Greilich_Science06,Greilich_Science07} in which the nuclear state was effectively narrowed by driving an ensemble of dots with a sequence of optical pulses.
We also identify the dominant processes which lead to SE decoherence at these low fields, which are the hf-mediated flip-flops of nuclei having different Zeeman energies, i.e.~nuclei of different isotopes of Ga and As. This crucial role of hetero-nuclear character of the bath in III-V materials at low B fields has not been appreciated before. 
At high magnetic fields this leads to a small oscillation superimposed on the SE signal due to the spectral diffusion, while at low fields it leads to a practically complete decay of the SE in less than a microsecond in large GaAs dots.
This result might provide an explanation for the recent measurements of the single-spin SE decay on time-scale of a microsecond in gated GaAs quantum dots.\cite{Koppens_PRL08}

Our work, developed in details in this article, brings the theory of quantum dephasing of the localized electron spin qubit due to its hf interaction with the nuclear spin to the same level of sophistication and quantitative depth that the current theories of spectral diffusion due to the dipolar nuclear flip-flops have.  As such, we can carry out quantitative comparison with experiments performed at low magnetic fields, where the hf coupling would dominate over the spectral diffusion process.  It is therefore gratifying that our theory gives reasonable quantitative agreement with the existing experimental data on spin dephasing down to a field as low as 10 mT.  More experimental data would be required for any further improvement in the theory.

% ORGANIZATION:
The paper is organized in the following way. In Sec.~\ref{sec:model} we describe the full Hamiltonian of the electron spin and the nuclear bath, and derive the effective pure dephasing Hamiltonian containing a hierarchy of multi-spin hf-mediated interactions. In Sec.~\ref{sec:W} we present a general formal framework for calculation of pure dephasing under the action of a sequence of ideal $\pi$ pulses. The main theoretical development is presented in Sec.~\ref{sec:rings}, where we identify the most important terms (corresponding to the ring diagrams) in the perturbation series for the decoherence function in the case of long-range interaction between the nuclear spins, and we show how they can be resummed.
This approach is then used in Secs.~\ref{sec:FID} and \ref{sec:SE} to calculate the decoherence in single-spin (narrowed state) FID and SE experiments, respectively. 
Our  FID calculation is compared with other recent theories in Sec.~\ref{sec:FID_previous}, and in Sec.~\ref{sec:FID_Dicke} we show that the ring diagram approach to FID agrees with a result obtained in a completely different fashion in the special case of uniform coupling of the nuclei to the central spin. In Sec.~\ref{sec:FID_results} we present our results for FID in GaAs and InGaAs quantum dots. Sec.~\ref{sec:SE} contains the theory and results for SE decay in low $B$ fields, when the interaction between the nuclei of different species gives the dominant contribution to SE decoherence.
The theoretical predictions for multi-pulse dynamical decoupling sequences in the same magnetic field regime are given in Sec.~\ref{sec:DD}.
In Sec.~\ref{sec:higher} we discuss the corrections to the lowest-order hf-mediated interaction and argue that our approach is well controlled on experimentally relevant time-scale as long as $\delta \! \ll \! 1$ ($B \! \gg \! 3$ mT in GaAs dots with $N\! \sim \! 10^{6}$ nuclei). 

Before proceeding further, let us establish the nomenclature. By the short-time regime will mean $t \! \ll \! N/\mathcal{A}$ (which corresponds to $t \! \ll \! 20$ $\mu$s in GaAs with $N \! =\! 10^{6}$ nuclei), while $t\! \gg \! N/\mathcal{A}$ is the long-time regime. By high magnetic field $B$ we will mean $B$ such that $\mathcal{A}/\Omega \! < \! 1$  (corresponding to $B \! \gtrsim \! 3$ T in large GaAs dots), while low $B$ will mean that $\mathcal{A}/\Omega \! > \! 1$, but $\delta\! = \! \mathcal{A}/\Omega\sqrt{N} \! \ll \! 1$. The regime of very low $B$ corresponding to $\delta \! \gtrsim \! 1$ remains outside of the realm of validity of the theory presented below.

%%%%%%%%%
%%% TABLE WITH ALL THE ENERGY SCALES 
%%%%%%%%%
\begin{table}
\begin{ruledtabular}
\begin{tabular}{ccc}
$\Omega/ g_{\text{eff}}$                				& $8.8\cdot10^{10}$ s$^{-1}$ 	&  $0.68$ K\\
$\omega_{i}$  ($^{69}$Ga)       		& $-6.42\cdot 10^{7}$ s$^{-1}$  					&  $-0.494$ mK\\
$\omega_{i}$  ($^{71}$Ga)       		& $-8.16\cdot 10^{7}$ s$^{-1}$  					&  $-0.628$ mK\\
$\omega_{i}$ ($^{75}$As)       		& $-4.58\cdot 10^{7}$ s$^{-1}$  					&  $-0.353$ mK\\
$\omega_{i}$  ($^{113}$In)       			  & $-5.85\cdot 10^{7}$ s$^{-1}$                                        &  $-0.450$ mK\\
$\omega_{i}$ ($^{115}$In)       			  & $-5.86\cdot 10^{7}$ s$^{-1}$                                        &  $-0.451$ mK\\
$\mathcal{A}$ $(^{69}$Ga)	       	& $5.47\cdot 10^{10}$ s$^{-1}$  				&  $0.421$ K\\
$\mathcal{A}$ $(^{71}$Ga) 		& $6.99\cdot 10^{10}$ s$^{-1}$  				&  $0.538$ K\\
$\mathcal{A}$ $(^{75}$As) 		& $6.53\cdot 10^{10}$ s$^{-1}$  				&  $0.503$ K\\
$\mathcal{A}$ $(^{113}$In)	       	  & $8.51\cdot 10^{10}$ s$^{-1}$                                       &  $0.655$ K\\
$\mathcal{A}$ $(^{115}$In) 		         & $8.53\cdot 10^{10}$ s$^{-1}$                                       &  $0.657$ K\\
$B_{ij}$						& $\sim 10^{2}$ s$^{-1}$  						&  $\sim 1$ nK \\
\end{tabular}
\end{ruledtabular}
\caption{Typical values of energies in the Hamiltonian in units of angular frequency ($\hbar$$=$$1$) and temperature ($k_{B}\! = \! 1$). The Zeeman energies $\Omega$ and $\omega_{i}$ are given for $B \! =\! 1$ T. The dipolar interaction $B_{ij}$ is given for two nuclei separated by $1$ nm distance. GaAs parameters are taken from Ref.~\onlinecite{Paget_PRB77} (see also Ref.~\onlinecite{Schliemann_JPC03}).
The values of $\mathcal{A}_{\alpha}$ for In are taken from Ref.~\onlinecite{Liu_NJP07}.
} \label{tab:energies} 
\end{table}
% CONVERSION FACTORS:
% 1 K = 0.086 meV
% 1/s / 2pi = Hz
% 1 meV = 1.52 1e12 1/s

%%%%%%%
%% THE  MODEL
%%%%%%%
\section{The Model} \label{sec:model}
The full Hamiltonian of the electron spin interacting with the  the nuclear bath is given by 
\beq
\hat{H} = \hat{H}_{\text{Z}} + \hat{H}_{\text{dip}}  + \hat{H}_{\text{hf}}  \,\,  ,
\eeq
with the terms corresponding to Zeeman, dipolar, and hf interactions, respectively. In order to describe the nuclear system in III-V semiconductors we have to account for the existence of nuclei of different species (isotopes), which we label by $\alpha$. To the lattice sites $i$ we assign randomly the nuclear species $\alpha[i]$, and in the subsequent calculations the average over possible nuclei on a given site will be taken implicitly, using the weights $n_{\alpha}$ denoting the number of $\alpha$ species nuclei per unit cell.  Nuclei of different $\alpha$  have different gyromagnetic factors, resulting in different Zeeman splittings, hf couplings, and dipolar couplings. Thus we write Zeeman and dipolar energies as 
\begin{eqnarray}
\hat{H}_{\text{Z}} & = & \Omega S^{z} + \sum_{i} \omega_{\alpha [i]} \hat{J}^{z}_{i} \,\, ,\label{eq:H_Z} \\  
\hat{H}_{\text{dip}} & = & \sum_{i\neq j} b_{ij} ( \hat{J}^{+}_{i}\hat{J}^{-}_{j} -2  \hat{J}^{z}_{i}\hat{J}^{z}_{j} ) \,\, , \label{eq:H_dip} 
\end{eqnarray}
and the hf interaction is given in Eq.~(\ref{eq:hf}). 
In these equations $\Omega$ is the electron's spin Zeeman splitting, $\omega_{\alpha [i]}$ are nuclear Zeeman splittings for nuclei at the $i$-th site, and $b_{ij}$ is the secular dipolar interaction between the nuclei, with $\alpha[i] \! = \! \alpha[j]$ implied.\cite{footnote_secular} 
The secular approximation to the dipolar Hamiltonian is justified by the fact that $b_{ij} \! \ll \! \omega_{i},\omega_{j}$ and $b_{ij} \! \ll \! \omega_{\alpha\beta}\! = \! \omega_{\alpha}-\omega_{\beta}$ for $\alpha\neq\beta$ for magnetic fields of interest here (see Table \ref{tab:energies}). 
The hf couplings $A_{i}$ in Eq.~(\ref{eq:hf}) are proportional to the total hf interaction energy $\mathcal{A}_{\alpha[i]}$ of the nuclear species $\alpha[i]$ at site $i$, and the square of the envelope function of the electron $f_{i} \equiv |\Psi(\mathbf{r}_{i}) |^{2}$:
\beq
A_{i} = \mathcal{A}_{\alpha[i]}f_{i} \,\, ,
\eeq
with the envelope normalization given by:
\beq
\int_{V} |\Psi(\mathbf{r})|^{2} d\mathbf{r}= \nu_{0}  \,\,\,\,\, \Leftrightarrow  \,\,\,\,\, \sum_{i} f_{i} = n_{c} \,\, ,  \label{eq:normalization}
\eeq
where $V$ is the volume of the quantum dot,  $\nu_{0}$ is the volume of the primitive unit cell ($\nu_{0} \! = \! a_{0}^3/4$ with $a_{0}$ being the lattice constant in zinc-blende or diamond structrure), and $n_{c}$ is the number of nuclei in the primitive unit cell (equal to $2$ in GaAs). 
In terms of microscopic parameters, we have\cite{Paget_PRB77}
\beq
\mathcal{A}_{\alpha} = \frac{2}{3}\mu_{0} \hbar^{2} \gamma_{S} \gamma_{J\alpha} |u_{\alpha}|^{2} \,\, ,
\eeq
with $\mu_{0}$ the vacuum permeability, $\gamma_{S}$ and $\gamma_{J\alpha}$ are the electron and nuclear spin gyromagnetic factors, respectively, and $u_{\alpha}$ is the amplitude of the periodic part of the Bloch function at the position of the nucleus of $\alpha$ species (the normalization is $\int_{\nu_{0}} |u(\mathbf{r})|^{2}d\mathbf{r} \! = \! 1$). The values of $\mathcal{A}_{\alpha}$ for Ga, As, and In are given in Table \ref{tab:energies}.
%The latter is often written as $|u_{\alpha}|^{2} \! = \! \eta_{\alpha}/\nu_{0}$, introducing the dimensionless parameter $\eta_{\alpha}$ (see Table \ref{tab:A}).

% DEFINITION OF N :
It will be convenient later to use the following definition of $N$, the number of nuclei interacting appreciably with the electron:
\beq
N \equiv  \frac{\sum_{i}f_{i}} {\sum_{i} f^{2}_{i}} = \frac{\int |\Psi(\mathbf{r})|^{2} d\mathbf{r}} {\int |\Psi(\mathbf{r})|^{4} d\mathbf{r}} \,\, , \label{eq:N}
\eeq
from which we get that when summing over the nuclei of one species $\alpha$ we have
\beq
\sum_{i\in \alpha} A^{2}_{i} = n_{\alpha} \mathcal{A}^{2}_{\alpha} / N  \,\, ,
\eeq
where we have used the fact that $\sum_{i\in \alpha}$ can be replaced by $(n_{\alpha}/n_{c})\sum_{i}$ when dealing with a large dot (i.e.~when there are enough nuclei for the distribution of $A_{i}$ to be self-averaging). In the same way we get the total hf interaction energy
\beq
\mathcal{A} \equiv \sum_{i}A_{i} = \sum_{\alpha}n_{\alpha}\mathcal{A}_{\alpha} \,\, ,
\eeq
which, using the parameters from Table \ref{tab:energies}, gives $\mathcal{A} \! = \! 83 \, \mu$eV in GaAs. Using the effective g-factor of $g_{\text{eff}} \! = \! -0.44$ in GaAs we get that $\mathcal{A}/\Omega \! = \! 1$ for magnetic field $B \! = \! 3.25$ T. 
On the other hand, the rms of the Overhauser field in an unpolarized thermal ensemble of nuclei is $\mathcal{A}/\sqrt{N}$, which for $N \! = \! 10^{6}$ corresponds to magnetic field of about $3$ mT. Thus, in the following by high fields we will mean $B \! > \! 3$ T and by low fields we mean $B \! < \! 3$ T (but larger than $\sim \! 10$ mT so that $\delta$ is indeed a small parameter).

%%%%%%%
%% DISTRIBUTION OF Ai
%%%%%%%
\subsection{Distribution of hyperfine couplings $A_{i}$}
In the following we will assume that the envelope wave-function of the electron in the dot is 
\beq
\Psi(\mathbf{r}) =  \sqrt{\frac{2\nu_{0}}{z_{0}\pi L^{2}} } \cos\left( \frac{\pi z}{z_{0}} \right ) \exp\left ( -\frac{x^{2}+y^{2}}{2L^{2}} \right ) \,\, ,  \label{eq:wave_function_cos}
\eeq
with the thickness of the dot $z_{0}$ and the Fock-Darwin radius $L$. For this choice of the envelope, from Eq.~(\ref{eq:N}) we get that $N \! = \! 4z_{0}\pi L^2 / 3\nu_{0}$. This number is of the same order of magnitude as $N_{\text{m}} \! = \! z_{0}\pi L^2/2\nu_{0}$ defined by 
\beq
\max_{i} \, A_{i} \equiv \mathcal{A}/N_{\text{m}} \,\, ,  \label{eq:Nm}
\eeq 
so that whenever we encounter a sharp inequality involving $A_{i}$, we can replace $A_{i}$ in it with $\mathcal{A}/N$.

For systems with a large number of nuclei we can make a continuum approximation for the distribution of $A_{i}$. We write for any function $F(A_{i})$
\beq
\sum_{i\in \alpha} F(A_{i}) = n_{\alpha} \int_{0}^{\infty} F(A)\rho_{\alpha}(A) dA \,\, ,  \label{eq:sum_integral}
\eeq 
where the ``density of states'' of the hf couplings is 
\beq
\rho_{\alpha}(A) = \frac{1}{\nu_{0}}  \int_{V} \delta\left(A - \mathcal{A}_{\alpha} | \Psi(\mathbf{r}) |^{2} \right) d^{3}r \,\, , \label{eq:rho}
\eeq
so that $\rho(A)dA$ is the number of \emph{unit cells} in which the hf interaction energy is in the range $[A,A+dA]$. For  the electron wave-function from Eq.~(\ref{eq:wave_function_cos}) we get
\beq
\rho_{\alpha}(A) = \frac{3}{2\pi} \frac{N}{A} \arccos\sqrt{\frac{3NA}{8\mathcal{A}_{\alpha}}} \,\,  \Theta\left(\frac{8}{3} \frac{\mathcal{A}_{\alpha}}{N} - A \right)  \,\, , \label{eq:rho_cos}
\eeq
where $\Theta(x)$ is the Heaviside step function.d

%%%%%%%
%% EFFECTIVE HAMILTONIAN
%%%%%%%
\subsection{Effective Hamiltonian}  \label{sec:Heff}
We write the hf interaction as 
% \sum_{i} A_{i} \left[  \hat{J}^{z}_{i} \hat{S}^{z} +\frac{1}{2}  ( \hat{S}^{+}\hat{J}^{-}_{i} +  \hat{S}^{-}\hat{J}^{+}_{i} )  \right] =
\beq
\hat{H}_{\text{hf}} =  \sum_{i}A_{i} \hat{J}^{z}_{i} \hat{S}^{z} + \hat{V}_{\text{sf}} \,\, ,
\eeq
where $\Vsf = \frac{1}{2} \sum_{i} A_{i} ( \hat{S}^{+}\hat{J}^{-}_{i} +  \hat{S}^{-}\hat{J}^{+}_{i} ) $
is the spin-flip part. In a finite magnetic field this interaction was shown\cite{Coish_PRB04,Shenvi_scaling_PRB05,Zhang_PRB06} to lead to relaxation of the $S^{z}$ component of the electron spin by a small quantity $\delta^{2} \! = \! \mathcal{A}^2/N\Omega^{2}$. In a large magnetic field when $\Omega \! \gg \! \mathcal{A}$, the direct contribution from $\Vsf$ can be neglected when considering the electron spin dephasing.\cite{Shenvi_bounds_PRB05} However, while the real $S\! - \!J$ spin flip is forbidden, higher order virtual processes in which the electron spin is flipped multiple times while having the same initial and final state, are still allowed. The existence of such processes is clearly brought to light by performing a canonical transformation removing $\Vsf$ from the Hamiltonian. Such a Schrieffer-Wolff type transformation leads to an appearance of effective hf-mediated interactions between the nuclei.\cite{Shenvi_scaling_PRB05,Yao_PRB06,Coish_PRB08,Cywinski_PRL09,Klauser_PRB08} 
We give a more detailed derivation of this transformation in Appendix \ref{app:canonical}. Here let us mention that in deriving the effective interactions we have assumed that $\Omega$ is much larger than the rms Overhauser shift in the nuclear state under considration. Since we are interested in thermal and unpolarized states of nuclei, this means $\Omega \! \gg \! \mathcal{A}/\sqrt{N}$ (i.e. $\delta \! \ll \! 1$). Under this transformation the Hamiltonian changes to $\tilde{H} \! =\! e^{-\mathcal{\hat{S}}} \hat{H} e^{\mathcal{\hat{S}}}$ and the states are transformed to $\tilde{\ket{\phi}}\! = e^{-\mathcal{\hat{S}}} \ket{\phi}$ with a carefully chosen unitary operator $e^{-\mathcal{\hat{S}}}$.
An important observation is the fact that the transformation of the states can be shown\cite{Yao_PRB06} to lead to a partial decay (visibility loss) of the coherence occuring on time-scale $t \! < \! N/\mathcal{A}$ and having the amplitude of $\delta^{2}$, in qualitative agreement with results from the GME theory.\cite{Coish_PRB04}
This leads to the conjecture that the ``effective Hamiltonian'' approach is physically sensible \emph{at least up to some time-scale} as long as $\delta \! \ll \! 1$. We elaborate on this issue in Sec.~\ref{sec:higher}, where we show that at the experimentally relevant\cite{Koppens_PRL08} time-scale (less than $\sim \! 10$ $\mu$s in GaAs with $N\! = \! 10^{6}$ nuclei) this condition is in fact sufficient for our approach to be well-controlled. Since this discussion involves time-dependence, it has to be preceded by the development of a theory of decoherence due to various terms in $\tilde{H}$, which  is given in Secs.~\ref{sec:W}-\ref{sec:SE}.

In all of the following we will assume $\delta \! \ll \! 1$ and neglect the transformation of states accompanying the transformation of $\hat{H}$. 
Using the method described in Appendix \ref{app:canonical} we arrive at the effective pure dephasing Hamiltonian
\beq 
\tilde{H} = \HH_{Z} + \sum_{i} A_{i} \hat{S}^{z}\hat{J}^{z}_{i} +\HH_{\text{dip}} + \sum_{n=2}^{\infty} \tilde{H}^{(n)} \,\, ,  \label{eq:Heff}
\eeq
where $\tilde{H}^{(n)}$ represent nuclear interaction terms generated in $n$-th order with respect to $\Vsf$. The  $\tilde{H}^{(n)}$ terms can contain only $\hat{S}^{z}$ operator of the electron spin, and thus in Eq.~(\ref{eq:Heff}) we have an effective \emph{pure dephasing} Hamiltonian.

In the second order we get a previously known result\cite{Shenvi_scaling_PRB05,Yao_PRB06,Coish_PRB08,Klauser_PRB08}
\begin{eqnarray}
\tilde{H}^{(2)} & =  & -\sum_{i} \frac{A^{2}_{i}}{4\Omega} \hat{J}^{z}_{i}  + \hat{S}^{z}\sum_{i} \frac{A^{2}_{i}}{2\Omega} \Big(\hat{J}_{i}^{2} - (\hat{J}^{z}_{i})^{2} \Big)  + \nonumber \\ 
& & + \hat{S}^{z}\sum_{i\neq j} \frac{A_{i} A_{j}}{2 \Omega} \hat{J}^{+}_{i}\hat{J}^{-}_{j}  \,\, . \label{eq:H2}  
\end{eqnarray}
The first term is a very small correction to Zeeman energy of nuclei, corresponding to the renormalization $\omega_{i}$ in the original Hamiltonian by $\Delta\omega_{i}^{(2)} \! = \! -A^{2}_{i}/4\Omega$, which is a negligible correction for $B$ fields above a millitesla. 
The second term is also rather small, and for $J \! = \! 1/2$ it amounts to a renormalization\cite{Coish_PRB08} of electron spin splitting: $\Delta\Omega^{(2)} \! \approx \! \mathcal{A}^{2}/4N\Omega$. The third term is the hf-mediated two-spin (2s) flip-flop interaction, which will be the focus of most of our attention:
\beq
\tilde{H}^{(2)}_{2s} = 2\hat{S}^{z} \sum_{i\neq j} B_{ij} \hat{J}^{+}_{i}\hat{J}^{-}_{j} \,\, ,  \label{eq:H2_2s}
\eeq
where $B_{ij} \! = \! A_{i}A_{j}/4\Omega$.
Since the nuclear interaction in this term is multiplied by electron $\hat{S}^{z}$ operator, we will call terms of this kind the $S^{z}$-conditioned ones. The crucial characteristic of $\tilde{H}^{(2)}$ is that it describes a \emph{non-local} interaction: we have of the order of $N^{2}$ pairs of nuclei coupled by appreciable interaction constants $\sim \mathcal{A}^2/N^{2}\Omega$. Thus, even if the typical interactions energy of any given pair of nuclei is small, one has to remember that this weakness of couplings can be overcome by the number of coupled pairs of nuclei.

The full expression for purely hyperfine contribution to $\tilde{H}^{(3)}$ is reproduced in Appendix \ref{app:canonical}. In this order we also obtain terms due to mixing of hf with dipolar interactions and nuclear Zeeman terms. These turn out to be small corrections to the pre-existing terms or interactions derived from $\Vsf$ only. Involving the dipolar interactions makes also the effective coupling more local, while the most important interactions are the ``maximally nonlocal'' ones, such as the three-spin (3s) term
\beq
\tilde{H}^{(3)}_{3s} = -\hat{S}^{z} \sum_{i\neq j\neq k} \frac{A_{i}A_{j}A_{k}}{2\Omega^{2}} \hat{J}^{+}_{i} \hat{J}^{-}_{j} \hat{J}^{z}_{k} \,\, . \label{eq:H3_3s}
\eeq
Another important feature in the third order is the appearance of the $S^{z}$-independent hf interaction:
\beq
\tilde{H}^{(3)}_{2s} = -\sum_{i\neq j} \frac{A_{i}A_{j}(A_{i}+A_{j})}{16\Omega^{2}}  \hat{J}^{+}_{i} \hat{J}^{-}_{j} \,\, .  \label{eq:H3_2s}
\eeq
The $S^{z}$-conditioned two-spin interactions in $\tilde{H}^{(3)}$ are small corrections to the flip-flop interaction from $\tilde{H}^{(2)}_{2s}$, since their coupling constants are scaled down by $\mathcal{A}/{N\Omega} \! \ll \! 1$.

In the following Sections we will concentrate on the $S^{z}$-conditioned two-spin interaction from $\tilde{H}^{(2)}_{2s}$ considered previously in Refs.~\onlinecite{Yao_PRB06,Liu_NJP07,Yao_PRL07,Coish_PRB08}, and on $S^{z}$-independent $\tilde{H}^{(3)}_{2s}$ terms. The influence of the higher-order multi-spin terms (such as $\tilde{H}^{(3)}_{3s}$) will be discussed in Sec.~\ref{sec:higher}.

%%%%%%%
%% W UNDER PULSES - KELDYSH CONTOUR, INTERACTION PICTRE
%%%%%%%
\section{General approach to decoherence under pulse sequences}  \label{sec:W}
We write the pure dephasing effective Hamiltonian as
\beq
\tilde{H} = \HH_{0} + \hat{V} = \HH_{\text{Zn}} + 2S^{z}\HH_{A} + 2S^{z}\hat{V}_{1} + \hat{V}_{2} \,\, ,  \label{eq:HV1V2}
\eeq
in which $\HH_{\text{Zn}}$ is the Zeeman energy of nuclei, $2S^{z}\HH_{A}$ is their diagonal hf coupling to the electron spin ($\HH_{A} = \sum_{i} \frac{A_{i}}{2}J^{z}_{i}$), and all the interactions between the nuclei are divided into $S^{z}$-conditioned ones in $\hat{V}_{1}$ and $S^{z}$-independent in $\hat{V}_{2}$ (which could include also the dipolar interactions). The electron Zeeman energy has been dropped from $\tilde{H}$ since it contributes only a simple phase factor of $\Omega t$ to the FID signal and completely cancels out from the expressions for SE decay.
We consider the evolution operator $\hat{U}(t)$ of the whole system under the action of a series of ideal ($\delta$-shaped) $\pi$ pulses applied to the electron spin. For pulses corresponding to rotations by angle $\pi$ about the $\hat{x}$ axis we have
\begin{equation}
\hat{U}(t) = (-i)^{n} \, e^{-i\hat{H}\tau_{n+1}} \hat{\sigma}_{x} e^{-i\hat{H}\tau_{n}} \, ... \, \hat{\sigma}_{x} e^{-i\hat{H}\tau_{1}} 
\end{equation}
with $n$ being the number of applied pulses, $\tau_{i}$ being time delays between the pulses, and the total evolution time $t \! = \! \sum_{i=1}^{n+1} \tau_{i}$. For example, the spin echo corresponds to $n\!=\!1$ and $\tau_{1} \! = \! \tau_{2} \! = \! t/2$.

The coherence of the electron spin is given by the off-diagonal element of its reduced density matrix, given by
\beq
\rho_{+-}(t) = \bra{+} \text{Tr}_{\text{B}} \hat{\rho}(t) \ket{-} \,\, ,  \label{eq:rhopm}
\eeq
where $\ket{\pm}$ are the eigenstates of $\hat{S}^{z}$, $\hat{\rho}$ is the density matrix of the whole system (electron$+$nuclei), and $\text{Tr}_{\text{B}}$ denotes tracing over the bath degrees of freedom. We assume the factorizable initial density matrix given by $\hat{\rho}(0) \! = \! \hat{\rho}_{\text{S}}(0) \hat{\rho}_{\text{B}}(0)$. 
With this  we arrive at the following expression\cite{Witzel_PRL07,Yao_PRL07,Lutchyn_PRB08} for the decoherence function $W(t)$ for even $n$:
\beq
W(t) \equiv \frac{\rho_{+-}(t)}{\rho_{+-}(0)}   =   \left \langle \hat{U}^{\dagger} _{-}(t) \hat{U}_{+}(t)  \right \rangle  \,\, , \label{eq:WU}
\eeq
and for odd $n$ there is $\rho_{-+}(0)$ in the denominator. This definition ensures $W(0) \! =\! 1$.
In Eq.~(\ref{eq:WU}) we have $\mean{...} \! = \! \text{Tr}_{\text{B}} \{\hat{\rho}_{\text{B}}(0)...\} $ denoting the average over the states of the nuclei, and $\hat{U}_{\pm}$ are evolution operators of the nuclear system conditioned on the state of the electron spin:\cite{Lutchyn_PRB08}
\begin{eqnarray}
\hat{U}(t)\ket{\pm}  & = & \ket{\pm} \hat{U}_{\pm}(t) \,\,\,\,\,\, \text{for} \,\,  n \,\,  \text{even} \,\, , \\
\hat{U}(t)\ket{\pm}  & = & \ket{\mp} \hat{U}_{\mp}(t) \,\,\,\,\,\, \text{for} \,\, n \,\, \text{odd} \,\, .
\end{eqnarray}
For example, we have for  FID
\beq
W_{\text{FID}}(t) =   \left \langle e^{i\tilde{H}_{-}t} e^{-i\tilde{H}_{+}t} \right \rangle  \label{eq:WU_FID}
\eeq
with $\tilde{H}_{\pm} \! = \! \hat{H}_{\text{Zn}}  \pm \HH_{A}  \pm \hat{V}_{1} + \hat{V}_{2}$, and for SE we have
\begin{eqnarray}
\hat{U}^{SE}_{\pm} & = & e^{-i\tilde{H}_{\pm}t/2} e^{-i\tilde{H}_{\mp}t/2} \,\, ,  \label{eq:U_SE}
\end{eqnarray}
and the expression for $W_{\text{SE}}(t)$ is obtained by plugging these into Eq.~(\ref{eq:WU}). 

Note that in Ref.~\onlinecite{Cywinski_PRL09} we defined $W(t) \! \equiv \! 2| \rho_{+-}(t)|$. As we discuss below, the two definitions are equivalent for SE and other dynamical decoupling sequences, but for FID the earlier choice meant discarding of the phase dynamics of $\rho_{+-}$, which will turn out to be nontrivial. Experimentally, the current definition corresponds to measuring $\mean{\hat{S}^{x}(t)}$ or $\mean{\hat{S}^{y}(t)}$, while the previous one corresponded to measurement of $\mean{\hat{S}^{x}(t)}^{2}  \!+ \!  \mean{\hat{S}^{y}(t)}^{2}$. The latter would require averaging over a series of measurements in which either $x$ or $y$ components of the spin are measured. 

%%%%%%%%
%%% W(t) ON KELDYSH CONTOUR 
%%%%%%%%
\subsection{Decoherence function on a contour}  \label{sec:contour}
The  evolution under the influence of pulses can be re-expressed by making $\HH_{A}$ and $\hat{V}_{1}$ time-dependent. Let us define the filter function in the time domain $f(t;\tau)$:
\beq
f(t;\tau) \equiv  (-1)^{n} \sum_{k=0}^{n} (-1)^{k} \Theta(\tau-t_{k}) \Theta(t_{k+1}-\tau) \,\, ,
\eeq
where $t_{k}$ with $k\! = \!1...n$ are the times at which the pulses are applied, $t_{0}\! = \! 0$ and $t_{n+1} \! = \! t$,
so that $f(t;\tau)$ is nonzero only for $\tau \in[0,t]$. Any pulse sequence can be encoded by this function,\cite{deSousa_review,Cywinski_PRB08} e.g.~the CPMG sequence\cite{Haeberlen} corresponds to $t_{k} \! = \! t(k-1/2)/n$. Using this function we can write the closed formula for $\hat{U}_{\pm}$ as
\beq
\hat{U}_{\pm}(t)  =  \mathcal{T} \exp\left[-i\int_{0}^{t}(\hat{H}_{\text{Zn}} + \hat{V_{2}} \pm f(t;\tau)[\HH_{A} +\hat{V}_{1} ] ) d\tau \right] \,\, ,
\eeq
where $\mathcal{T}$ is the time ordering operator. Then we transform into the interaction picture with respect to $\hat{V}(t) = \hat{V}_{2} \pm f(t;t')\hat{V}_{1}$. The evolution operator in the interaction picture, $S_{\pm}(t)$, is defined by
\begin{eqnarray}
\hat{U}_{\pm}(t) & = & \left( \mathcal{T} \exp\left[-i\int_{0}^{t} \hat{H}^{\pm}_{0}(\tau) d\tau \right] \right) S_{\pm}(t) \,\, , \\
& \equiv  & \hat{U}^{\pm}_{0}(t) S_{\pm}(t)
\end{eqnarray}
where $\hat{H}^{\pm}_{0}(t') \! = \! \HH_{\text{Zn}}  \pm f(t;\tau) \HH_{A} $, so that the time-ordering is actually unnecessary and in expression for $\hat{U}^{\pm}_{0}(t)$ and it is given by
\beq
\hat{U}^{\pm}_{0}(t) = \exp \left[-i\HH_{\text{Zn}}t \mp i\HH_{A}\int_{0}^{t}f(t;\tau)d\tau \right] \,\, .
\eeq
From this we get
\begin{eqnarray}
\hat{S}_{\pm}(t) & = & \mathcal{T}  \exp\left[-i\int_{0}^{t}  \mathcal{\hat{V}}_{\pm}(\tau)   d\tau \right] \,\, , \\
\hat{S}^{\dagger}_{\pm}(t) & = & \bar{\mathcal{T}}  \exp\left[i\int_{0}^{t}  \mathcal{\hat{V}}_{\pm}(\tau)   d\tau \right] \,\, ,
\end{eqnarray}
where $\bar{\mathcal{T}}$ is the time anti-ordering operator and 
\beq
\hat{\mathcal{V}}_{\pm}(\tau) = \left(\hat{U}^{\pm}_{0}(\tau)\right)^{\dagger} \hat{V} \, \hat{U}^{\pm}_{0}(\tau) \,\, .
\eeq
Then, for any balanced pulse sequence (defined by $\int f(t;\tau)d\tau \! = \! 0$, which \emph{does not} hold for FID, see Sec.~\ref{sec:W_FID}), the decoherence function becomes:
\beq
W(t) =  \left\langle \hat{S}^{\dagger}_{-}(t)\hat{S}_{+}(t) \right\rangle \,\, .  \label{eq:W_SS}
\eeq

\begin{figure}[t]
\centering
\includegraphics[height=5cm]{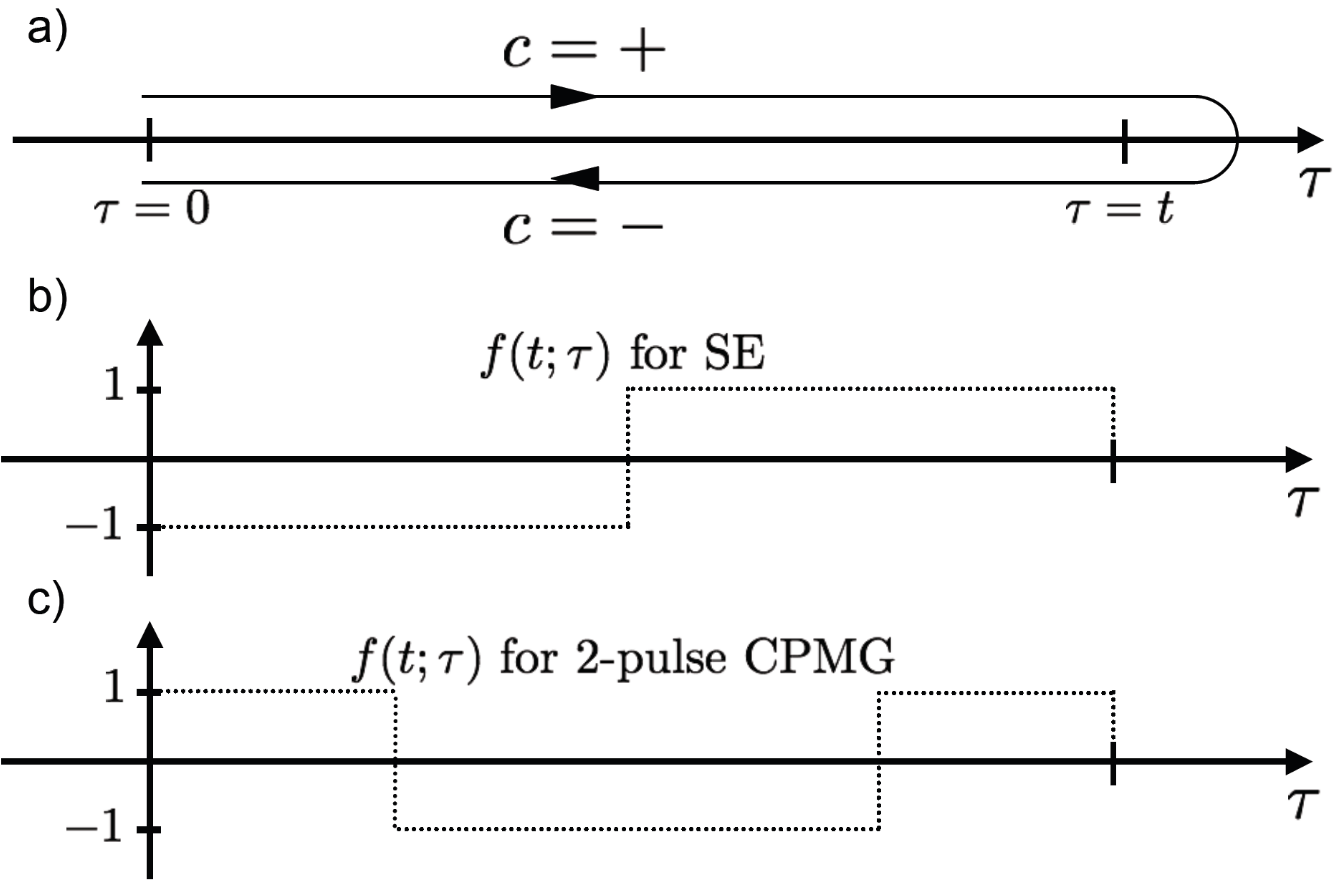}
  \caption{ a) The closed loop contour along which the operators $\hat{\mathcal{V}}(\tau,c)$ are ordered. b)
 The plot of the time-domain filter function $f(t;\tau)$ for the Spin Echo sequence. c) The same for 2-pulse CPMG sequence. } \label{fig:contour}
\end{figure}

In Eqs.~(\ref{eq:WU}) and (\ref{eq:W_SS}) one can see that $W(t)$ is the average of the evolution of the bath first from $0$ to time $t$, and then back (under a different Hamiltonian) from $t$ to $0$. This structure calls for use of a closed-time-loop technique well-established for nonequilibrium problems in many-body theory.\cite{Rammer} Such a Keldysh-like structure has been used previously in decoherence problems.\cite{Grishin_PRB05,Saikin_PRB07,Yang_CCE_PRB08,Lutchyn_PRB08} 
Introducing the notion of ordering the operators on the Keldysh contour  (see Fig.~\ref{fig:contour}) allows us to write $W(t)$ in a more compact way:
\beq
W(t) =  \left\langle \mathcal{T}_{C} \exp \left[ -i\int_{C} \hat{\mathcal{V}}(\tau_{c}) d\tau_{c} \right] \right\rangle \,\, . \label{eq:W_contour}
\eeq
In the above equation $\mathcal{T}_{C}$ denotes the contour-ordering of operators,\cite{Rammer} and $\tau_{c} \! = \! (\tau,c)$ is the time variable on the contour, with $c\!=\! + (-)$ on the upper (lower) branch of the contour and $\tau$ being the corresponding real time value. 
The contour integration of a function $F(\tau_{c}) \! = \! F(\tau,c)$ is given by
\beq
\int_{C} F(\tau,c) d\tau_{c} = \int_{0}^{\infty} F(\tau,1) d\tau + \int_{\infty}^{0} F(\tau,-1) d\tau \,\, .  \label{eq:contour_int}
\eeq
In terms of the free-evolution operator on the contour:
\beq
\hat{\mathcal{U}}_{0}(\tau_{c}) = \exp \left[- i\HH_{\text{Zn}}\tau - ic\int_{0}^{\tau}f(t;t')\HH_{A} dt'  \right] \,\, , 
\eeq
we have the interaction operators
\begin{eqnarray}
\hat{\mathcal{V}}_{1}(\tau_{c}) & = & c f(t;\tau) \, \hat{\mathcal{U}}^{\dagger}_{0} (\tau_{c}) \, \hat{V}_{1} \, \hat{\mathcal{U}}_{0}(\tau_{c})    \,\, ,\\
\hat{\mathcal{V}}_{2}(\tau_{c}) & = & \hat{\mathcal{U}}^{\dagger}_{0}(\tau_{c}) \, \hat{V}_{2} \, \hat{\mathcal{U}}_{0}(\tau_{c})  \,\, .
\end{eqnarray}
Note that the sign of the $\hat{\mathcal{V}}_{1}$ interaction is opposite on two branches of the contour. The $\hat{\mathcal{V}}_{1,2}$ interactions are given by the same formulas as the Schroedringer picture $\hat{V}_{1,2}$, only with the spin operators replaced by their branch- and time-dependent counterparts:
\begin{eqnarray}
\hat{J}^{\pm}_{j}(\tau_{c}) & = & \hat{J}^{\pm}_{j} \exp\left [ \pm i \omega_{j}\tau \pm i c \int_{0}^{\tau}f(t;t')\frac{A_{j}}{2} dt' \right ] \,\, , ~ ~
%& \equiv & \hat{J}^{\pm}_{j} d^{\pm}_{j}(\tau_{c}) 
 \label{eq:JpJm}
\end{eqnarray}
%where we have defined the $\tau_{c}$-dependent function $d^{\pm}_{j}(\tau_{c})$, 
and $\hat{J}^{z}(\tau_{c}) \! = \! \hat{J}^{z}$.  

%%%%%%
%%% AVERAGING OVER THE NUCLEI
%%%%%%
\subsection{Averaging over the nuclear ensemble}  \label{sec:averaging}
The averaging over the nuclear ensemble is performed here using a thermal density matrix of nuclei $\hat{\rho}_{\text{B}} \! = \! e^{-\beta\HH_{\text{B}}}/Z$, where $Z$ is the partition function. Due to the smallness of the energy scale associated with nuclear Hamiltonian $\HH_{\text{B}} \! = \! \hat{H}_{Zn}+\hat{H}_{\text{dip}}$, at the experimentally achievable temperatures we approximate it by $\hat{\rho}_{\text{B}} \approx 1/Z$. Thus, we treat the case of the ``infinite temperature'' unpolarized nuclear bath. Then, the average in Eq.~(\ref{eq:W_SS}) for balanced pulse sequences is a real number and there are only even orders of $\hat{V}$ in the perturbation expansion of $W(t)$ (see Ref.~\onlinecite{Witzel_PRB06}).

An important feature of the thermal nuclear bath is that it is uncorrelated, i.e.~the average of a product of nuclear operators corresponding to disjoint sets of nuclei factorizes:
\begin{eqnarray}
 & &\mean{F_{1}(\mathbf{J}_{i_{1}},...,\mathbf{J}_{i_{n_{1}}})\, F_{2}(\mathbf{J}_{j_{1}},...,\mathbf{J}_{j_{n_{2}}}) } =  \nonumber\\
 & & \mean{ F_{1}(\mathbf{J}_{i_{1}},...,\mathbf{J}_{i_{n_{1}}}) } \mean{ F_{2}(\mathbf{J}_{j_{1}},...,\mathbf{J}_{j_{n_{2}}}) } \,\, , \label{eq:factorization}
\end{eqnarray}
where $\{i_{1},...,i_{n_{1}}\}$ and  $\{j_{1},...,j_{n_{2}}\}$ are two non-overlapping sets of nuclei. Below we will extensively use this factorization in the case when among the averaged operators there are only two ($\hat{J}^{+}_{i}$ and $\hat{J}^{-}_{i}$) corresponding  to a certain nuclear spin $i$. In such a case we have
\beq
\mean{... \hat{J}^{\pm}_{i} ... \hat{J}^{\mp}_{i} ...} = \mean{\hat{J}^{+}_{i}\hat{J}^{-}_{i}} \mean{...} \,\, ,  \label{eq:JpJm_av}
\eeq
where 
\beq
\mean{\hat{J}^{+}_{i}\hat{J}^{-}_{i}}  = \frac{2}{3} J_{i}(J_{i}+1) \equiv a_{i} \,\, ,  \label{eq:adef}
\eeq
for a nuclear spin $J_{i}$. In the most relevant case here of GaAs quantum dots, we have $J_{i} \! = \! 3/2$ and $a_{i} \! = \! 5/2$ for all the nuclei, while for nuclei of Indium we have $J_{i} \! = \! 9/2$ and $a_{i} \! = \! 33/2$.

%%%%%%
%%% FID - NARROWED STATES ETC...
%%%%%%
\subsection{Decoherence function for Free Induction Decay}  \label{sec:W_FID}
In the case of FID, the diagonal hf interaction $\HH_{A}$ in $\hat{U}^{\pm}_{0}$  does not cancel out from $W(t)$, and instead of Eq.~(\ref{eq:W_SS}) we obtain
\beq
W_{\text{FID}}(t) =  \left\langle \hat{S}^{\dagger}_{-}(t)\, e^{-2i\HH_{A}t}  \, \hat{S}_{+}(t) \right\rangle \,\, . 
\eeq
Using the ``infinite temperature'' nuclear thermal ensemble we arrive at\cite{Yao_PRB06}
\beq
W_{\text{FID}}(t) =   \frac{1}{Z} \sum_{\mathcal{J}} e^{-i\Omega_{\mathcal{J}}t} \bra{\mathcal{J}} \hat{S}_{+}(t)  \hat{S}^{\dagger}_{-}(t) \ket{\mathcal{J}}   \,\, ,   \label{eq:W_FID_sum}
\eeq
where $\ket{\mathcal{J}} \! = \! \otimes_{i} \ket{j^{z}_{i}}$ are the product states of the eigenvectors of $J^{z}_{i}$ operators of all the nuclear spins, and $\Omega_{\mathcal{J}}$ is the Overhauser shift of the electron frequency for the nuclear state $\ket{\mathcal{J}}$:
\beq
\Omega_{\mathcal{J}} =  \bra{\mathcal{J}} 2 \hat{H}_{A} \ket{\mathcal{J}}  = \sum_{i} A_{i}j^{z}_{i} \,\, .
\eeq
The form of the decoherence function in Eq.~(\ref{eq:W_FID_sum}) is quite different from the one in Eq.~(\ref{eq:W_SS}). However, it has been shown in Refs.~\onlinecite{Yao_PRB06} and \onlinecite{Liu_NJP07} that the matrix element in this equation for practical purposes does not depend on the state $\ket{\mathcal{J}}$, at least as long as we consider the  ``typical'' states with polarization close to zero. 
This follows from the fact that there are many nuclei in the bath making it self-averaging. Let us note that this is not the case for dilute baths, such as the dipolarly coupled spin bath interacting with the NV center in diamond, where different spatial realization of the bath can lead to very different FID signals.\cite{Dobrovitski_PRB08}

Consequently, in the dense hf-coupled bath we make an approximation of replacing the $\bra{\mathcal{J}}...\ket{\mathcal{J}}$ matrix element with its ensemble average, obtaining
\begin{eqnarray}
W_{\text{FID}}(t) & \approx & \sum_{\mathcal{J}} e^{-i\Omega_{\mathcal{J}}t} \times \left\langle \hat{S}^{\dagger}_{-}(t)  \hat{S}_{+}(t)  \right\rangle  \,\, ,  \\
& \equiv & W^{\text{i}}_{\text{FID}}(t) \times W^{\text{s}}_{\text{FID}} \,\, .  \label{eq:FID_factorization}
\end{eqnarray}
In the above Equation, $W^{\text{i}}_{\text{FID}}$ gives the fast FID decay due to  inhomogeneous broadening:\cite{Khaetskii_PRL02,Merkulov_PRB02,Liu_NJP07} $W^{\text{i}}_{\text{FID}} \sim \exp[-(t/T^{*}_{2})^2]$ 
where $T^{*}_{2}\!  \approx \! \sqrt{N}\mathcal{A} \! \sim \! 10$ ns, while $W^{\text{s}}_{\text{FID}}$ is the decoherence function \emph{ for a single spin}.  The latter quantity, which has the same form as Eq.~(\ref{eq:W_SS}), is of main interest here. 

The measurement of the single spin FID is in principle possible by devising an experimental procedure which leads to ``narrowing'' of the distribution of the nuclear states,\cite{Klauser_PRB06,Giedke_PRA06,Stepanenko_PRL06} with only states having the same $\Omega_{\mathcal{J}}$ contributing in the fully narrowed situation. There has recently been experimental progress in performing such experiments,\cite{Greilich_Science06,Greilich_Science07,Reilly_Science08} with an estimate of the single-spin $T_{2}$ time of about $3$ $\mu$s in InGaAs dots.\cite{Greilich_Science06}

The above-described approach for calculation of $W^{\text{s}}_{\text{FID}}$ is equivalent to calculating the FID evolution for a single ``typical'' $\ket{\mathcal{J}}$ state, which was employed in Refs.~\onlinecite{Coish_PRB04} and \onlinecite{Coish_PRB08}. While in most of the following we will concentrate on the calculation in which $W^{\text{s}}_{\text{FID}}$ is calculated by full ensemble average, with $\Omega_{\mathcal{J}}$ shifts factored out according to Eq.~(\ref{eq:FID_factorization}), in Sec.~\ref{sec:FID_Dicke} we will compare this approach with another method in which we explicitly worked with a narrowed nuclear state.

%%%%%%%
%% RING DIAGRAMS
%%%%%%%
\section{Ring diagram expansion of decoherence function}  \label{sec:rings}
The theoretical framework presented so far is very general. The calculation of the decoherence function given in Eq.~(\ref{eq:W_SS}) has been achieved for the nuclei interacting via dipolar interactions using different versions of exponential resummation of the perturbation series for $W(t)$.\cite{Witzel_PRB05,Witzel_PRB06,Yao_PRB06,Liu_NJP07,Saikin_PRB07,Witzel_PRB08} A similar approach has been used for the case of nuclei interacting via the lowest-order $S^{z}$-conditioned hf-mediated interaction,\cite{Yao_PRB06,Liu_NJP07,Saikin_PRB07} but only the lowest-order terms in the cumulant expansion have been retained in these calculations. In this Section we show how to resum all the leading (in terms of $1/N$ expansion) terms in the perturbation series for $W(t)$. We start with an illustrative example calculation of the lowest-order terms in Sec.~\ref{sec:W2}, give the full theory of the resummation of  ring diagrams in Sec.~\ref{sec:resummation}, discuss its relation to previous works in Sec.~\ref{sec:previous}, and finally present a numerically efficient method of calculating the sum of all the ring diagrams in Sec.~\ref{sec:T}.

In this Section and in Secs.~\ref{sec:FID} and \ref{sec:SE} we will concentrate on two-spin hf-mediated interactions, such as the leading $S^{z}$-conditioned interaction  in $\tilde{H}$ given in Eq.~(\ref{eq:H2_2s}) and the $S^{z}$-independent term from Eq.~(\ref{eq:H3_2s}). The role of multi-spin interactions will be discussed in Sec.~\ref{sec:higher}.

%%%%%%
%%% SECOND ORDER W
%%%%%%
\subsection{Lowest order contributions to $W(t)$}  \label{sec:W2}
For concreteness, let us use the two-spin interaction from Eq.~(\ref{eq:H2_2s}). The corresponding interaction on the contour is given by
\begin{eqnarray}
\hat{\mathcal{V}}(\tau_{c}) & = & c f(t;\tau) \sum_{i\neq j} B_{ij} J^{+}_{i}(\tau_{c})J^{-}_{j}(\tau_{c}) \,\, , \label{eq:VSz} \\
& \equiv & \sum_{i\neq j} \mathcal{V}_{ij}(\tau_{c}) \hat{J}^{+}_{i}\hat{J}^{-}_{j} \,\, , \label{eq:Vkl}
\end{eqnarray}
with the contour-time dependence of the ladder operators given in Eq.~(\ref{eq:JpJm}) and with $B_{ij} \! = \! A_{i}A_{j}/4\Omega$. For the two-spin interaction from Eq.~(\ref{eq:H3_2s}) the calculation is analogous, only there is no $c  f(t;\tau)$ factor in Eq.~(\ref{eq:VSz}) and the couplings $B_{ij}$ are different.

Perturbation expansion of $W(t)$ defined in Eq.~(\ref{eq:W_contour}) leads to the series $W \! = \! 1+ \sum_{k=2} W^{(k)}$, with $W^{(k)} \! \sim \! \mean{\mathcal{V}^{k}}$. In fact, for SE and any balanced sequence of pulses only the even orders are non-zero and only for FID we need to consider odd $k$. 
In order for the product of $\mathcal{\hat{V}}$ not to average to zero, all the raising operators $\hat{J}^{+}_{i}$ have to be paired up with lowering operators $\hat{J}^{-}_{i}$. 
In the second order we have
\begin{eqnarray}
W^{(2)}(t) & = & -\frac{1}{2}\int_{C}d\tau_{c_{1}}\int_{C}d\tau_{c_2}  \mean{ \mathcal{T}_{C} \hat{\mathcal{V}}(\tau_{c_1}) \hat{\mathcal{V}}(\tau_{c_2}) } \,\, ,~~~
\end{eqnarray}
where  we used the notation $\tau_{c_{i}} \! = \! (\tau_{i},c_{i})$. The averaged expression is
\begin{eqnarray}
\mean{ \mathcal{T}_{C} \hat{\mathcal{V}}(\tau_{c_1})\hat{\mathcal{V}}(\tau_{c_2}) }  & = & \sum_{k\neq l}  \sum_{m\neq n} \mathcal{V}_{kl}  (\tau_{c_1}) \mathcal{V}_{mn}(\tau_{c_2})  \nonumber \\
& & \!\!\!\!\!\!\!\!\!\!\!\!\!\!\!\!\!\!\!\!\!\!\!\!\!\!\!\!\!\!  \mean{  \mathcal{T}_{C} \hat{J}^{+}_{k} \hat{J}^{-}_{l} \hat{J}^{+}_{m}  \hat{J}^{-}_{n} } \,\, ,
%& & \!\!\!\!\!\!\!\!\!\!\!\!\!\!\!\!\!\!\!\!\!\!\!\!\!\!\!\!\!\!  \mean{  \mathcal{T}_{C} \hat{J}^{+}_{k}(\tau_{c_1}) \hat{J}^{-}_{l}(\tau_{c_1}) \hat{J}^{+}_{m}(\tau_{c_2}) \hat{J}^{-}_{n}(\tau_{c_2}) } \,\, , 
\end{eqnarray}
and since $k\! \neq \! l$ and $m \! \neq \! n$ there is only one way of contracting the operators:
\begin{equation}
 \begin{array}{l}
\contraction{J^{+}_{k}}{J^{-}_{l}}{}{J^{+}_{m}} 
\contraction[2ex] {} { J^{+}_{k}} {J^{-}_{l}J^{+}_{m}} {J^{-}_{n}}
\mean{ J^{+}_{k}J^{-}_{l}J^{+}_{m}J^{-}_{n} } = \delta_{kn}\delta_{lm} \mean{ J^{+}_{k}J^{-}_{l}J^{+}_{l}J^{-}_{k} } \,\, .
\end{array}
\end{equation}
Let us note now that due to $k \neq l$ constraint,  the ladder operators effectively commute under the sign of the average: the only non-zero commutator is $[\hat{J}^{+}_{i},\hat{J}^{-}_{i}] \! = \! 2\hat{J}^{z}_{i}$ with $i\! = \! k,l$, and while moving the operators around we can create at most one $J^{z}_{i}$ operator. The average of an expression with a single $\hat{J}^{z}_{i}$ being the only operator pertaining to the $i$-th spin can be factored as in Eq.~(\ref{eq:factorization}) and one can see then that the whole expression is equal to zero since $\mean{\hat{J}^{z}_{i}}\! = \! 0$ in the unpolarized bath. Therefore the spin operators are effectively commuting under the average,  and we can remove the time-ordering arriving at
\beq
W^{(2)}(t) =  -\frac{1}{2} \sum_{k,l} T_{kl}(t) T_{lk}(t) =  -\frac{1}{2} \text{Tr} \mathbf{T}^{2}  \label{eq:W2}
\eeq
where we have defined the matrix $T_{kl}(t)$:
\begin{eqnarray}
T_{kl}(t) & \equiv & (1-\delta_{kl})  \sqrt{a_{k}a_{l}} \int_{C} d\tau_{c} \mathcal{V}_{kl}(\tau_{c})  \,\, , \label{eq:Tdef}
\end{eqnarray}
with $a_{k}$ defined in Eq.~(\ref{eq:adef}) and $ \mathcal{V}_{kl}$ defined in Eq.~(\ref{eq:Vkl}).
This quantity can be readily evaluated for any two-spin interaction and any pulse sequence. 

% EXPLANATION OF DIAGRAMS ADDED:
An expression from Eq.~(\ref{eq:W2}) can be represented graphically by a diagram shown in Fig.~\ref{fig:rings}a, where the nuclei are represented by the dots, and the interactions integrated over the time along the contour (i.e.~the elements of the $T_{kl}$ matrix) are represented by lines. In the corresponding analytical expression each dot contributes a sum over all the $N$ nuclei, with the restriction that only terms with all the indices corresponding to distinct nuclei are summed over in the whole expression. Physically the diagram in Fig.~\ref{fig:rings}a describes a process in which a pair of nuclei flips twice during their evolution along the ``time'' contour and comes back to the original state (so that the whole process survives the averaging procedure). Similarly, the higher order diagrams in Fig.~\ref{fig:rings} correspond to processes of propagation of a nuclear spin flip along a closed loop involving more than two nuclei. While the probability of a single such event quickly decreases with the number $k$ of the nuclei involved, the number of such events grows as $N^{k}$. The ring diagrams represent the sums over all such events.

In an analogous manner one can calculate the third order term $W^{(3)}(t)$: there we have two possible contractions, but similarly to the case of $W^{(2)}$ we have then a sum over three distinct nuclear indices, and the ladder operators are again effectively commuting. The final result is
\beq
W^{(3)}(t)  =  \frac{(-i)^{3}}{3!}  2 \sum_{k\neq l \neq m} T_{kl}(t)T_{lm}(t)T_{mk}(t) \,\, ,  \label{eq:W3} 
\eeq
which is represented graphically in Fig.~\ref{fig:rings}b.

However, the higher orders of perturbative expansion of $W(t)$ do not have the same simple structure. Below we discuss the approximation under which all the most important contributions to $W^{(k)}(t)$ can be easily calculated in any order $k$.

%%%%%%%%
%%% EXPONENTIAL RESUMMATION OF W(t)
%%%%%%%%
\subsection{Exponential resummation of $W(t)$}  \label{sec:resummation}
When calculated exactly, the higher order terms in expansion of $W(t)$ are quite cumbersome, since the diagrammatic expansion\cite{Saikin_PRB07,Yang_CCE_PRB08} is complicated by the lack of simple Wick's theorem for spin operators, and in order to account for all the terms one has to introduce more complicated diagram construction rules than the ones given for diagrams corresponding to $W^{(2)}$ and  $W^{(3)}$.    
Now we will describe an approximate way of resumming of the perturbation series for decoherence due to the non-local interactions between the nuclei. 

The $W^{(k)}$ term has the following structure:
\beq
W^{(k)} \sim \sum_{i_{1}\neq j_{1}} ... \!\! \sum_{i_{k}\neq j_{k}} \mean{\mathcal{T}_{C} J^{+}_{i_{1}}J^{-}_{j_{1}} \,\, ... \,\, J^{+}_{i_{k}}J^{-}_{j_{k}} } \,\, ,
\eeq
so that it is a sum of $[N(N-1)]^{k}$ averages  of time-ordered products of operators. These averages are non-zero only when all the $J^{+}$ and $J^{-}$ operators are paired up, so that there are at most $\sim \! N^{k}$ terms contributing to the sum. 
From these sums over $k$ nuclear indices we now take only the terms in which none of the indices is repeated.
Assuming that all the terms are of the same order of magnitude, this amounts to at most $1/N$ error compared to the exact expression. 
The simplification following from this step is tremendous: now, as in the cases considered in Sec.~\ref{sec:W2}, the operators effectively commute under the average and we can get rid of the time-ordering. Let us stress that in this approximation we are neglecting all the terms which correspond to diagrams having more interaction lines than distinct nuclei. Rules for constructing these subleading (for long-range interaction) diagrams are complicated,\cite{Saikin_PRB07,Yang_CCE_PRB08} but fortunately we do not need to use them here.

\begin{figure}[t]
\centering
\includegraphics[width=0.9\linewidth]{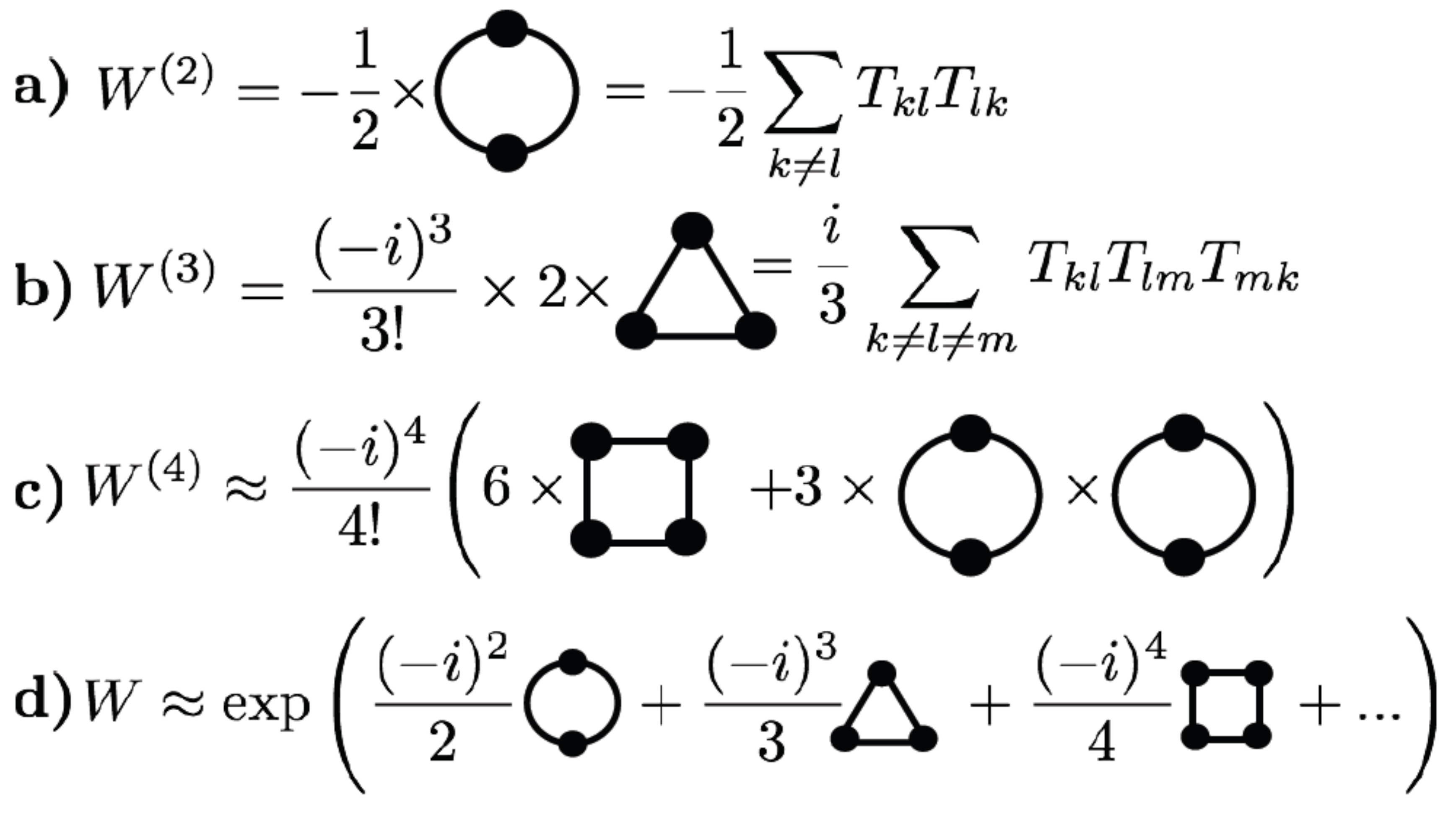}
  \caption{(a) and (b) Graphical representation of exact $W^{(2)}$ from Eq.~(\ref{eq:W2}) and $W^{(3)}$  from Eq.~(\ref{eq:W3}) . (c) Ring diagram approximations for $W^{(4)}$ from Eq.~(\ref{eq:W4}). (d) Linked cluster resummation of $W(t)$.}  \label{fig:rings}
\end{figure}

Now, only the way in which the spin operators are paired up matters. In the $4$th order we have two distinct patterns of contractions, see Fig.~\ref{fig:rings}c. There are $6$ ways of contracting the operators in such a way that the interactions $B_{ij}$ are connecting all the nuclei in a closed cycle (a four-nuclei ring), and $3$ ways of forming two disjoint cycles ($2$-nuclei rings). The sum is over  $4$ distinct nuclear indices, but we will invoke the $1/N$ expansion again and allow for the indices to repeat if they appear in disjoint rings connected by interaction, leading us to 
\beq
W^{(4)}(t)  \approx  \frac{(-i)^{4}}{4!} ( 6 R_{4} + 3R_{2}^{2} ) \,\, , \label{eq:W4}
\eeq
and in a similar way we get for the higher order terms:
\begin{eqnarray}
W^{(5)}(t) & \approx &  \frac{(-i)^{5}}{5!} ( 4! R_{5} + 20 R_{3}R_{2} ) \,\, ,\\
W^{(6)}(t) & \approx & \frac{(-i)^{6}}{6!} ( 5! R_{6} + 90 R_{4}R_{2} + 5!! R^{3}_{2} ) \,\, , \text{etc}
\end{eqnarray}
where we have defined the \emph{ring diagrams} $R_{k}$:
\beq
R_{k}(t) = \sum_{i_{1}\neq i_{2} \neq ... \neq i_{k}} T_{i_{1}i_{2}}(t) ... T_{i_{k}i_{1}}(t) \approx \text{Tr}\mathbf{T}(t)^{k} \,\, .  \label{eq:R}
\eeq
The sum on the left-hand side of the above expression is restricted to distinct nuclei (i.e.~$i_{n}\! \neq \! i_{m}$ for each $n$ and $m$). Consequently, the approximation in the rightmost part of the above formula again introduces a  $1/N$ error. 

The combinatorics of generating ring-diagram approximations to higher-order terms in expansion of $W(t)$ quickly becomes tedious. For example, in $W^{(2k)}$  the number of contractions leading to $R_{2}^{k}$ terms is given by $(2k-1)!!$, and the number of $R_{k}$ terms in $W^{(k)}$ is $(k-1)!$. However, for further progress we only need to know the coefficient in front of $R_{k}$ contribution to $W^{(k)}$, since this is the only \emph{connected} term: all the others are products of multiple rings.

Here we use a very general linked-cluster (or cumulant expansion) theorem,\cite{Kubo_JPSJ62,Negele} according to which for a quantity $W(t)$ being an average of a generalized exponent (e.g.~an ordered exponential of operators), the logarithm of $W(t)$ is given by the sum of all the terms in expansion of $W(t)$ which are irreducible, i.e.~cannot be written as products of averages. In the diagrammatic language this means that $\ln W(t)$ is the sum of all the linked diagrams with combinatorial prefactors with which they appear in expansion of $W(t)$. A short proof of this theorem (along the lines of Ref.~\onlinecite{Negele}) is given in Appendix \ref{app:replica}. In $k$-th order of expansion we have then the linked term:
\beq
W^{(k)}_{\text{linked}} = \frac{(-i)^{k}}{k!} (k-1)! \, R_{k} = \frac{(-i)^{k}}{k}\, R_{k} 
\eeq
leading to the closed expression for the decoherence function
\beq
W(t) \approx \exp \left( \sum_{k=2}^{\infty} \frac{(-i)^{k}}{k} R_{k}(t) \right)  \,\, , \label{eq:W_R}
\eeq
where the $\approx$ sign reminds the fact that we have used $1/N$ approximation, and that we have replaced the upper summation limit (which is of the order of $N$) by infinity. The ring diagram terms $R_{k}(t)$ are given by Eq.~(\ref{eq:R}). The graphical representation of this equation is shown in Fig.~\ref{fig:rings}d.
For SE and other balanced pulse sequences we then have
\beq
W_{\text{even}} (t)\approx \exp \left( \sum_{k=1}^{\infty} \frac{(-1)^{k}}{2k} R_{2k}(t) \right)  \,\, , \label{eq:W_R_even}
\eeq
These equations are the central formal result of this paper.

%%%%%
%% PREVIOUS CLUSTER-THEORIES
%%%%%
\subsection{Relation to previous cluster-expansion theories of electron spin dephasing}  \label{sec:previous}
Before moving on to discussing efficient ways of evaluating Eq.~(\ref{eq:W_R}), let us outline the relation of this theory to the  approaches previously used to calculate the pure dephasing due to the nuclear bath.
Retaining only the $R_{2}$ term in the exponent corresponds to the PCA of Refs.~\onlinecite{Yao_PRB06,Liu_NJP07}. Note that since $W^{(2)} \! = \! -(1/2)R_{2}$ exactly, without invoking $1/N$ expansion, $R_{2}$ is also the first  term in the linked-cluster expansion for the dipolar flip-flop interaction between the nuclei, and in Refs.~\onlinecite{Yao_PRB06,Liu_NJP07,Yao_PRL07} it was used to calculate $W(t)$ for both hf-mediated and dipolar interactions.
It was noted in Ref.~\onlinecite{Saikin_PRB07} that for the long-range interaction the diagrams connecting cyclically the maximal number of the nuclei in a given order of expansion are the most important. These are the ring diagrams considered here.

Furthermore, we can relate Eq.~(\ref{eq:W_R}) to the real space cluster expansion of Refs.~\onlinecite{Witzel_PRB05,Witzel_PRB06,Witzel_PRB08} by noticing that $R_{2k}$ are related to the cluster contributions $v'_{2k}$ defined in Ref.~\onlinecite{Witzel_PRB06}:
\beq
\frac{(-1)^{k}}{2k}R_{2k} = \sum v'_{2k}  \,\, \text{in the 2k-th order in $\mathcal{V}$,}
\eeq
where the sum is over all the clusters having $2k$ nuclei. 
From the perspective of real-space cluster expansion of  Refs.~\onlinecite{Witzel_PRB05,Witzel_PRB06,Witzel_PRB08} the 
exponentiation in Eq.~(\ref{eq:W_R}) serves to account, with correct combinatorial factors,
for all products of contributions from disjoint clusters that must be
included in $W(t)$; the $1/n!$ factor of the exponential expansion
serves to compensate for the $n!$ permutations over the $n$ cluster
contributions in each product.  Errors in this approximation arise
from extraneous products among overlapping (non-disjoint) clusters.
Here, these correspond to the aforementioned $O(1/N)$ errors.  In
these previous papers the cluster contributions $v'_{2k}$ were
calculated for nuclei coupled by dipolar interactions, which are \emph{not} long-ranged in the meaning used here, i.e.~they do not couple equally all the $N$ nuclei. Thus $R_{2k}$ with $k\!> \! 1$ are not expected to be as good approximations to the exact cumulant of order $2k$ as in the case of the hf-mediated interactions. Indeed, for dipolar interactions the above-mentioned cluster overlap corrections are important for accuracy of higher-order terms in the exponent.\cite{Witzel_PRB06} These corrections are related to more complicated diagrams appearing in the linked-cluster expansion,\cite{Saikin_PRB07} and recently an approach in which they are avoided has been proposed for the case of small nuclear baths.\cite{Yang_CCE_PRB08,Yang_CCE_PRB09}

In the case of dipolar interactions the contribution of $R_{2}$ was giving a certain decoherence time-scale, and the contributions of larger clusters were negligible \emph{at this time-scale}. This can be traced back to the fact that the number of relevant terms, which involve close-by nuclei with large couplings, was of the same order of $N$ in all considered orders of expansion. In such a case the correct small parameter for the expansion is $bt$, with $b$ being the largest possible dipolar coupling, and consequently it is enough to consider only small clusters of a few nuclei. The situation is different for hf-mediated interaction: the number of terms to consider in $k$-th order is $N^{k}$, and at low magnetic fields (when the hf-mediated couplings are not very small) it is not \emph{a priori} clear that one can consider only the smallest clusters of nuclei. In fact, in the following we will show that at experimentally relevant magnetic fields and time-scales \emph{all} the ring diagrams are of the same order, and their resummation is necessary to obtain quantitatively correct results.

%%%%%%%
%%% T - MATRICES
%%%%%%%
\subsection{Calculation of all the ring diagrams using T-matrices}  \label{sec:T}
Eq.~(\ref{eq:R}) can be rewritten using the eigenvalues $\lambda_{l}(t)$ of the $T$-matrix as $R_{k} (t)= \sum_{l}^{N} \lambda_{l}^{k} (t)$. 
%\beq
%R_{k} (t)= \sum_{l}^{N} \lambda_{l}^{k} (t)\,\, .  \label{eq:R_lambda}
%\eeq
Then, from Eq.~(\ref{eq:W_R_even}) we get 
\begin{eqnarray}
&& \!\!\!\!\!\!\!\!\!\!\!\!\!\!\!\!\!\!\!\! W(t)_{\text{even}}  =  \prod_{l}^{N} \exp \left( -\frac{1}{2} \sum_{k=1}^{\infty} \frac{(-1)^{k+1}}{k} \lambda_{l}(t)^{2k} \right) \label{eq:Wsumlog} \\
&&\!\!\!\!\!\!\!\!\!\!\!\!\!\!\!\!\!\!\!\!  = \prod_{l}^{N} \exp \left( -\frac{1}{2}  \ln \left( 1+ \lambda_{l}(t)^{2} \right)   \right) =  \prod_{l}^{N} \frac{1}{\sqrt{1 + \lambda_{l}(t)^{2} } }\,\, ,  \label{eq:W_lambda} 
\end{eqnarray}
and in the analogous way the summation over the odd orders in Eq.~(\ref{eq:W_R}) gives
\beq
W(t)_{\text{odd}} = \prod_{l}^{N} \exp \Big( i[\lambda_{l}(t) - \arctan \lambda_{l}(t) ] \Big) \,\, ,  \label{eq:W_lambda_odd} 
\eeq
and for FID we have $W^{s}_{\text{FID}} \! = \! W_{\text{even}} \times W_{\text{odd}}$.

It remains to be shown that the calculation of $W(t)$ using Eqs.~(\ref{eq:R}) and (\ref{eq:W_R}) is feasible. The $T_{ij}$ matrix with $i$ and $j$ indices labeling the nuclei has the  dimension of $N\times N$, which is too large for direct calculation. One possible solution is to use the continuum approximation, and replace the sums over the nuclei by integrals over the densities of hf couplings $\rho(A)$ defined in Eq.~(\ref{eq:rho}). As for the errors, since we are effectively replacing the smallest possible $A_{ij} \! \sim \! \mathcal{A}/N^{2}$ with zero, we can expect these to arise at time $t \! > \! N^{2}/\mathcal{A}\! \sim \! 1 $ s, which is an irrelevantly long time-scale. However, a drawback of this method is that the main merit of the approach from Eq.~(\ref{eq:R}), which is being able to obtain \emph{all} the ring diagrams from a single calculation of the $T$-matrix, is lost.  Only under certain conditions we will be able to obtain formulas for all the $R_{k}$ from the multiple integrals in the continuum approximation. 

We can take full advantage of Eq.~(\ref{eq:R}) if we notice that $T_{kl}$ can be written as
\beq
T_{kl} = f_{kl} B(A_{k},A_{l}) \,\, ,
\eeq
where $B(A_{k},A_{l})$ is the coupling constant ( i.e.~$A_{1}A_{2}/4\Omega$ for the lowest-order $S^{z}$-conditioned interaction), and $f_{kl}$ depends only on  
\begin{eqnarray}
A_{kl} & \equiv & (A_{k}-A_{l})/2 \,\, , \\
\omega_{kl} & \equiv & \omega_{k} - \omega_{l} \,\, .
\end{eqnarray} 
Now, starting from the continuum approximation for $R_{k}$, we derive a formula just like Eq.~(\ref{eq:R}), but with an effective $\tilde{T}$ matrix replacing the original $T$.
We coarse-grain the distribution $\rho(A)$, dividing the relevant range of $A_{i}$ into $M_{A}$ slices, and deal with a $\tilde{T}$ matrix of size $M \! = \! M_{A}N_{J}$, where $N_{J}$ is the number of the nuclear species.
We take a discrete set of $A_{\alpha k}$ values for each species $\alpha$ with $A_{\alpha k} \! = \! k\Delta_{\alpha}$ with $ \Delta_{\alpha} \!= \! A^{\text{max}}_{\alpha}/M_{A}$. The coarse-grained matrix is defined as
\begin{eqnarray}
\tilde{T}_{\alpha k,\beta l} & = & f_{kl}  \Bigg( n_{\alpha} n_{\beta} \int_{A_{\alpha k}}^{A_{\alpha k}+\Delta_{\alpha}} \rho_{\alpha}(A_{1})   \int_{A_{\beta l}}^{A_{\beta l}+\Delta_{\beta}} \rho_{\beta}(A_{2}) \nonumber\\
& &   B^{2}(A_{1},A_{2}) \, dA_{1} dA_{2} \Bigg)^{1/2}  \,\, . \label{eq:T_coarse}
\end{eqnarray}
Note that in the $M_{A} \rightarrow \infty$ limit $\tilde{T}_{kl}$ does not approach the original $T$-matrix, but the  trace of the product of coarse-grained matrices approaches the continuum approximation to the expression involving the original $T_{ij}$. This definition of $\tilde{T}_{kl}$ is convenient because for $M_{A} \! =\! 1$ it is equivalent to expanding the continuum approximation expression to the lowest order in time (assuming $t \! \ll \! N/\mathcal{A}$).

While in the original $T$-matrix we had diagonal elements equal to zero, in the coarse-grained matrix we have $\tilde{T}_{kk} \! \neq \! 0$, because  these matrix elements represents the interactions between $\sim \! N/M_{A}$ nuclei, and removing them would correspond to an artificial lower bound on allowed $A_{ij}$. Similarly as in the case of continuum approximation, this approach is expected to lead to visible errors for $t \! > \!  NM_{A}/\mathcal{A}$. 
It will turn out that the calculation of $W(t)$ quickly converges at the time-scales of interest as we increase $M$ (i.e.~make the distribution of $A_{i}$ more fine-grained), and it is enough to use $M \! \ll \! N$.

%%%%%%%
%%%%%%%
%%  FID 
%%%%%%%
%%%%%%%
\section{Free Induction Decay} \label{sec:FID}
We consider now the FID experiment, characterized by $f(t;\tau) \!= \!1$ for $\tau \! \in [0,t]$ and $0$ otherwise.
For the lowest order $S^{z}$-conditioned interaction from Eq.~(\ref{eq:H2_2s}) we have
\begin{eqnarray}
T_{ij} & = & (1-\delta_{ij}) \sqrt{a_{i}a_{j}} \frac{A_{i}A_{j}}{4\Omega}  \frac{2}{ A^{2}_{ij} - \omega^{2}_{ij} }  \nonumber \\
& & \!\!\!\!\!\!\!\!\!\!\!\!\!\!\!\! \!\!\!\!\!\!  \left( e^{i\omega_{ij}t} \Big[ i\omega_{kl}\cos A_{ij}t+ A_{ij}\sin A_{ij}t \Big ] - i\omega_{ij} \right ) \,\, .  \label{eq:T2_FID}
\end{eqnarray}
The $T_{ij}$ matrix elements for higher-order two-spin interactions (e.g.~the $S^{z}$-independent pair interaction from Eq.~(\ref{eq:H3_2s})) are suppressed by powers of very small quantity $\mathcal{A}/N\Omega$.
% FULL EXPRESSION COMMENTED OUT:
%we have
%\begin{eqnarray}
%T_{ij} & = & (1-\delta_{ij}) \sqrt{a_{i}a_{j}} \frac{A_{i}A_{j}(A_{i}+A_{j})}{16\Omega^{2}}  \frac{2i}{ A^{2}_{ij} - \omega^{2}_{ij} }  \nonumber \\
%& & \!\!\!\!\!\!\!\! \left ( A_{ij}+e^{i\omega_{ij}t} \Big[ i\sin A_{ij}t - A_{ij}\cos A_{ij}t \Big] \right ) \,\, ,
%\end{eqnarray}
%which is smaller by a factor of $\mathcal{A}/N\Omega \! \ll \! 1$ from the $T_{ij}$ in Eq.~\ref{eq:T2_FID}.

When $\omega_{ij} \! \gg A_{ij}$ (which is true in  GaAs with $N \! \sim \! 10^{5}-10^{6}$ and $B \! \sim \! 0.1$T), the $T_{ij}$ from Eq.~(\ref{eq:T2_FID}) for $i$ and $j$ nuclei belonging to the same species are bigger than the hetero-nuclear matrix elements by a factor of the order of $\omega_{ij}/A_{ij}$. In this case, we can keep only the homo-nuclear terms in the $T$-matrix:
\beq
T_{ij} \approx (1-\delta_{ij})(1-\delta_{\alpha[i],\beta[j]}) a_{\alpha[i]} \frac{A_{i}A_{j}}{4\Omega} \frac{2\sin A_{ij}t }{A_{ij}} \,\, .\label{eq:Thom}
\eeq
% HETERO-NUCLEAR PART COMMENTED OUT:
%\beq
%T^{\text{het}}_{ij} \approx   \mean{J^{+}J^{-}} \frac{A_{i}A_{j}}{4\Omega} \frac{-2i\left( e^{i\omega_{ij}t} \cos A_{ij}t - 1\right)  }{\omega_{ij}} \,\, ,  \label{eq:Thet}
%\eeq
This amounts to treating $N_{J}$ nuclear systems as disconnected, and to factoring the ring diagrams $R_{2k}$ into three contributions: $ R_{k} \approx \sum_{\alpha=1}^{N_{J}} R^{\alpha}_{k}$.
%\beq
%R_{k} \approx \sum_{\alpha=1}^{N_{J}} R^{\alpha}_{k} \,\, .
%\eeq
From this the factorization of $W(t)$ follows:
\beq
W^{s}_{\text{FID}}(t) \approx \prod_{\alpha}   e^{-i\Delta\Omega^{(2)}_{\alpha}t}  W_{\text{FID}}^{s,\alpha}(t) \,\, ,  \label{eq:WFID_factorization}
\eeq
with $W_{\text{FID}}^{s,\alpha}(t)$ given by Eqs.~(\ref{eq:W_R}), $R_{k}$ calculated according to Eq.~(\ref{eq:R}) using the $T$-matrix from Eq.~(\ref{eq:Thom}). The electron frequency shift $\Delta\Omega^{(2)}_{\alpha}$ comes from the second term in Eq.~(\ref{eq:H2}). Strictly speaking this term is a constant only for $J\!= \! 1/2$, but we will treat it in mean-field approximation, i.e.~we will replace it with its expectation value with respect to the narrowed state $\ket{\mathcal{J}}$: 
\beq
\Delta\Omega^{(2)}_{\alpha} \approx n_{\alpha}a_{\alpha} \frac{\mathcal{A}^{2}_{\alpha}}{2N \Omega } \,\, .
\eeq

%%%%%
%%% FID AT SHORT TIMES
%%%%%
\subsection{Exact resummation for short times}  \label{sec:FID_short}
As a check of the calculation of $W(t)$ using the matrices following from discretization of the distribution of the hf couplings, let us consider the limit in which we can calculate all the $R_{k}$ analytically.
We restrict ourselves to the short-times ($A_{ij}t \approx \mathcal{A}t/N \! \ll \! 1$), which corresponds to $t \! \ll \! 20$ $\mu$s for a GaAs dot with $N\!= \! 10^{6}$. Then we can approximate the homo-nuclear $T$-matrix for nuclear species $\alpha$  by
\beq
T^{\alpha}_{ij} \approx  (1-\delta_{ij}) a_{\alpha} \frac{A_{i}A_{j}}{2\Omega} t \,\, ,  \label{eq:TFID_short}
\eeq
Using the continuum approximation we arrive at
\beq
R^{\alpha}_{k} =   \left( \frac{a_{\alpha}t}{2\Omega} n_{\alpha} \int A^{2} \rho_{\alpha}(A) dA \right)^{k} \,\, ,
\eeq
which, using the density of hf couplings for a lateral dot from Eq.~(\ref{eq:rho_cos}) gives us
\beq
R^{\alpha}_{k} =   \left( \frac{a_{\alpha}n_{\alpha}\mathcal{A}^{2}_{\alpha} t}{2N\Omega} \right)^{k}  \equiv (\eta_{\alpha}t)^{k}  \,\, .  \label{eq:R2_FID}
\eeq
Using the last equality we can resum the exponential series in Eqs.~(\ref{eq:W_R})  in the same manner in which we derived Eqs.~(\ref{eq:W_lambda}) and (\ref{eq:W_lambda_odd}). 
Using Eq.~(\ref{eq:WFID_factorization}), and noticing that $\Delta \Omega^{(2)}_{\alpha} \! = \! \eta_{\alpha}$, we arrive at the short-time expression for single-spin FID decoherence 
\beq
W^{s}_{\text{FID}}(t) = \prod_{\alpha} \frac{e^{-i\arctan \eta_{\alpha}t } }{\sqrt{1+ \eta^{2}_{\alpha}t^{2} } }  \,\, . \label{eq:WFID_short}
\eeq
The modulus of this expression was given in Ref.~\onlinecite{Cywinski_PRL09}.
Eq.~(\ref{eq:WFID_short}), together with the factorization formula in Eq.~(\ref{eq:WFID_factorization}), is the main analytical result pertaining to the FID decoherence in III-V quantum dots for short  times.
Taking advantage of the long-range nature of the hf-mediated interaction, we have resummed the whole perturbation series for $W(t)$ obtaining a simple analytical expression. The characteristic FID decoherence time, defined by $|W(T_{2})| \! = \! 1/e$, is
\beq
T_{2,\text{short}} \approx \frac{N\Omega}{\mathcal{A}^{2}} \,\, ,  \label{eq:T2}
\eeq
The same value was predicted within the PCA calculation,\cite{Yao_PRB06,Liu_NJP07} but the form of the decay was different there for $t \! \ll \! N/\mathcal{A}$. A similar characteristic time was also obtained using an equations of motion approach.\cite{Deng_PRB06,Deng_PRB08}

Let us stress that the $T_{2,\text{short}}$ gives a characteristic time of decay only if $T_{2,\text{short}}$ is shorter than $N/\mathcal{A}$, or equivalently when $\Omega/\mathcal{A} \! < \! 1$, i.e.~for low magnetic fields. We comment on the issue of the high-field and long-time decay in the following Section.

\subsection{Relation to previous work on FID}  \label{sec:FID_previous}
While our expression for $W(t)$ is given by Eq.~(\ref{eq:W_R}), the PCA result of Ref.~\onlinecite{Yao_PRB06} is given in our notation by
\beq
W_{\text{FID,PCA}}(t) = \exp \left( -\frac{1}{2} R^{\alpha}_{2}(t) \right) \,\, .
\eeq
The two formulas give indistinguishable results when $R_{2} \! \gg \! \sum_{k=3}^{\infty} R_{k}$. In the short-time regime this is fulfilled only when $R_{2} \! \ll \! 1$. Then, both approaches give 
\beq
W_{\text{FID}}(t) \approx  1- \frac{1}{2}\sum_{\alpha} \eta^{2}_{\alpha}t^{2} \approx \exp(- \frac{1}{2}\sum_{\alpha} \eta^{2}_{\alpha}t^{2}) \ \,\, ,  \label{eq:WFID_expt2}
\eeq
which holds when $\eta_{\alpha}t \! \ll \! 1$, or equivalently $t \! \ll \! (N/\mathcal{A})\cdot (\Omega/\mathcal{A})$. For $B \! <\! 1$ T in GaAs we have $\Omega/\mathcal{A} \! < \! 1$, and  this time-scale is shorter than the scale on which the Eq.~(\ref{eq:WFID_short}) is valid. Thus, at low fields most of the coherence decay is well described by Eq.~(\ref{eq:WFID_short}), with Gaussian decay being a good approximation only at very short times (when $W(t) \! \approx \! 1$). At low enough fields it should be possible to observe the regime in which $\eta_{\alpha}t \! \gg \! 1$, and the coherence signal is given by
\beq
W_{\text{FID}}(t) \approx \prod_{\alpha} \frac{e^{-i\pi/2} }{\eta_{\alpha}t} \,\, ,
\eeq 
which for GaAs means $W(t) \sim 1/t^{3}$. We stress that this formula holds when $t\! \gg \!  (\Omega/\mathcal{A})\cdot (N/\mathcal{A})$ and $t \! \ll \! N/\mathcal{A}$ (and for moderate magnetic field, since $\omega_{\alpha\beta} \! \gg \! A_{ij}$ is also required). At longer times and smaller $B$ fields we have to evaluate numerically the full expression for $W(t)$ instead of using Eq.~(\ref{eq:WFID_short}).

Interestingly, at long times ($t \! \gg \! N/\mathcal{A}$)  a very different form of $W(t)$ is obtained.
The expressions for ring diagrams are of the form
\begin{eqnarray}
R^{\alpha}_{k} & = & n^{k}_{\alpha} a^{k}_{\alpha} \int dA_{1} ... \int dA_{k} \rho(A_{1}) ... \rho(A_{k}) \, \frac{A^{2}_{1} ... A^{2}_{k}}{(2\Omega)^{k}} \nonumber\\
& & \times \frac{\sin A_{12}t}{A_{12}} \frac{\sin A_{23}t}{A_{23}} \, ... \,  \frac{\sin A_{k1}t}{A_{k1}} \,\, ,
\end{eqnarray}
and for the two-spin diagram we immediately get
\beq
R^{\alpha}_{2} = a^{2}_{\alpha}n^{2}_{\alpha} t^2 \int \rho_{\alpha}(A_{1})dA_{1} \int \rho_{\alpha}(A_{1})dA_{2} \frac{A^{2}_{1}A^{2}_{2}}{4\Omega^2} \text{sinc}^{2}A_{ij}t \,\, ,
\eeq
where $\text{sinc}\, x \! = \! \sin x/x$.
As discussed in Ref.~\onlinecite{Liu_NJP07}, in the $t \! \gg \! N/\mathcal{A}$ limit, when $\text{sinc}^{2}A_{ij}t \rightarrow \frac{\pi}{t}\delta(A_{ij})$, we obtain
\beq
R^{\alpha}_{2} \approx t \, \frac{\pi a^{2}_{\alpha}n^{2}_{\alpha}}{2\Omega^2} \int \rho^{2}_{\alpha}(A) A^{4} dA \equiv \frac{2t}{T^{\alpha}_{2,\text{long}}} \,\, .
\eeq
If $R_{2} \! \gg \! \sum_{k=3}^{\infty} R_{k}$ at these long times, then the coherence decay is given by
\beq
W(t \gg N/\mathcal{A}) \! \sim \! \prod_{\alpha}\exp(-t/T^{\alpha}_{2,\text{long}}) \,\, ,
\eeq
showing that the influence of the bath at long times can be treated in a Markovian approximation.
It is interesting to note that while the characteristic decay time $T_{2,\text{short}}$ at low fields depended on the electron wave function only through $N$ defined in Eq.~(\ref{eq:N}), the long-time decay constant $T^{\alpha}_{2,\text{long}}$ does depend\cite{Coish_PRB08} on the shape of $\Psi(\mathbf{r})$, specifically the distribution $\rho(A)$ of hf couplings.
For our wave-function from Eq.~(\ref{eq:wave_function_cos}) and the corresponding $\rho(A)$ from Eq.~(\ref{eq:rho}) we obtain
\beq
T^{\alpha}_{2,\text{long}} = \frac{3\pi N\Omega^{2}}{32n^{2}_{\alpha}a^{2}_{\alpha}\mathcal{A}^{3}_{\alpha}} \left( \frac{5\pi^2}{192}-\frac{17}{108} \right)^{-1} \,\, . \label{eq:T2_long}
\eeq

% InGaAs T2 figure  -  ADDED !!!
\begin{figure}[t]
\centering
\includegraphics[width=0.9\linewidth]{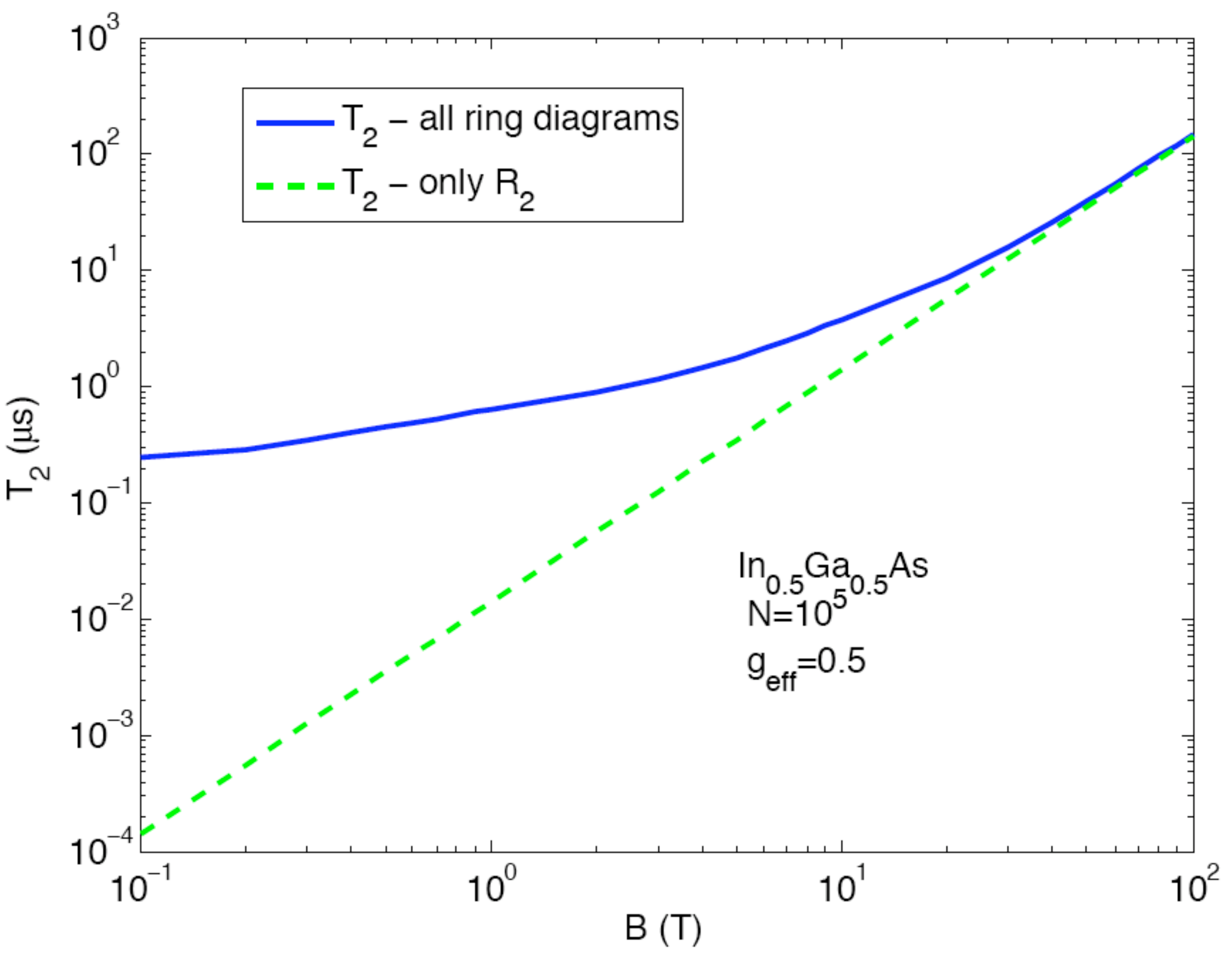}
  \caption{(Color online) Magnetic field dependence of the $T_{2}$ decay time characterizing the long-time exponential decay of coherence in FID experiment. The calculation is performed for In$_{0.5}$Ga$_{0.5}$As with $N\! = \!10^{4}$ and $g_{\text{eff}} \! = \! 0.5$: the solid line is the $T$-matrix resummation of all the ring diagrams ($M\! = \! 100$ was used), and the dashed line is the result of Eq.~(\ref{eq:T2_long}), with the total time given by $T_{2} \! = \! (\sum_{\alpha} T^{-1}_{\alpha})^{-1}$ }  \label{fig:T2_InGaAs}
\end{figure}

%%
%% COMPARISON WITH COISH'S PRB '08
%%
The same result for $|W(t)|$ at long times has been obtained within a very different approach (GME using hf-mediated interaction from Eq.~(\ref{eq:H2_2s})) in Ref.~\onlinecite{Coish_PRB08}. However, it was shown there that the result from Eq.~(\ref{eq:T2_long}) is valid only for $\mathcal{A}/\Omega \! \ll \! 1$. Indeed, we also find that the full $T$-matrix calculation (i.e.~resummation of all the ring diagrams) gives a long-time exponential decay with $T_{2,\text{long}}$ time given by Eq.~(\ref{eq:T2_long}) only when $\mathcal{A}/\Omega \! \ll \! 1$. At lower fields we still have $|W(t)| \! \sim \! \exp (-t/T_{2,\text{long}}^{\text{full}})$, but with $T_{2,\text{long}}^{\text{full}}$ larger than the value from Eq.~(\ref{eq:T2_long}).
In Fig.~\ref{fig:T2_InGaAs} we present the results of the calculation of magnetic field dependence of $T_{2,\text{long}}$ for In$_{0.5}$Ga$_{0.5}$As, where one can see that only for $B \! \gg \! 1$ T Eq.~(\ref{eq:T2_long}) gives a correct result.

Let us stress, that the exponential decay dominates the coherence dynamics at high fields ($\Omega \! > \! \mathcal{A}$), at which the initial decay of $|W(t)|$ in the short-time regime (discussed in the previous Section) is small, and most of decoherence occurs in the long-time regime. At low $B$ most of the decay occurs  in the short-time regime, and $T_{2,\text{long}}^{\text{full}}$ shown in Fig.~\ref{fig:T2_InGaAs} at these fields is the characteristic time of exponential decay which emerges as long times when the coherence is already small. However, as we discuss in Sections \ref{sec:FID_Dicke} and \ref{sec:higher}, we can safely argue that the ring diagram resummation approach is valid at low $B$ fields only for short times. Thus, the low-B and long time results, such as shown in Fig.~\ref{fig:T2_InGaAs} for $B\! < \! 3$ T, should be viewed with caution.

To summarize, at low $B$ our theory for FID  predicts exponential decay at long times when $|W(t)|$ is already very small, and the accuracy of this prediction has little practical bearing, while at high $B$ our results agree with other approaches.\cite{Coish_PRB08} For the purpose of quantum computation using spin qubits, which is the main motivation for our work, the long-time decay is of no significance whatsoever since one is only interested in the regime where quantum coherence is very high. However, for the purpose of establishing a closer connection between different theoretical approaches, the question of long-time and low-$B$ behavior of FID is an important one, and we leave it for future investigation. Finally, let us mention that based on comparison of ring diagram theory with exact simulation of a small system we believe that for SE our theory works well also for long times and relatively small magnetic fields.\cite{Slava}

%%%%%%%%%%%%%%%
%%% UNIFORM Ai COUPLINGS - DICKE
%%%%%%%%%%%%%%%
\subsection{FID for uniform hf couplings}   \label{sec:FID_Dicke}
We can test the ring diagram solution by applying it to the model system in which all the hf couplings $A_{i}$ are the same ($A_{i} \! = \! \mathcal{A}/N$), i.e.~the electron wave-function is assumed to be constant inside of the dot and zero outside.\cite{Taylor_PRL03,Melikidze_PRB04,Zhang_PRB06} 
We will also assume a homo-nuclear system with all the nuclei having the Zeeman splitting $\omega$, and take the nuclear spin $J \! = \! 1/2$. 
In such a case, we can rewrite the second order term in the effective Hamiltonian from Eq.~(\ref{eq:H2}) as
\begin{eqnarray}
\tilde{H}^{(2)} \!\!\! & = & \!\!\! (\Omega +\eta)\hat{S}^{z}  +  \omega'\hat{J}^{z} +   \frac{\mathcal{A}}{{N}} \hat{S}^{z}\hat{J}^{z}  + \hat{S}^{z} \frac{\mathcal{A}^{2}}{2\Omega N^{2}}  \sum_{i\neq j} \hat{J}^{+}_{i}J^{-}_{j} \nonumber\\
 & \!\!\!\!\!\!\!\!\!\!\!\!\!\!\!\! = &\!\!\!\!\!\!\!\!\!\! \Omega\hat{S}^{z} + \omega'\hat{J}^{z} +  \frac{\mathcal{A}}{{N}} \hat{S}^{z}\hat{J}^{z} + \hat{S}^{z} \frac{\mathcal{A}^{2}}{2\Omega N^{2}} ( \hat{J}^{2} - (\hat{J}^{z})^2 ) ~,~~\label{eq:H2_J}
\end{eqnarray}
where  $\eta \! = \! \mathcal{A}^{2}/4N\Omega$ and $\omega' \! = \! \omega - \mathcal{A}^{2}/4\Omega N^{2} $, and $\hat{J}$ ($\hat{J}^{z}$) are the operators of the total angular momentum of the nuclei (its projection on the $z$ axis). It is natural now to work in the basis of collective Dicke states $\ket{\beta j m }$, in which $j$ is the total spin of the nuclei, $m$ is the eigenvalue of $\hat{J}^{z}$, and $\beta$ is a permutation group quantum number.\cite{Arecchi_PRA72,Taylor_PRL03} 

In order to obtain the narrowed state FID decay we plug in the above Hamiltonian into Eq.~(\ref{eq:WU_FID}) in which the average corresponds to trace only over states with a fixed value of $m$. In the frame rotating with $\Omega$ frequency this gives us
\begin{eqnarray}
W_{m}(t) & = &\exp \Big[ -i(\Omega_{m} - \frac{\Omega_{m}^{2}}{2\Omega}   )t \Big] \frac{1}{Z_{m}} \sum_{j=|m|}^{N/2} n_{j} \nonumber\\
& & \times \exp \left( -i\frac{\mathcal{A}^2}{2\Omega N^{2}} j(j+1) t \right) \,\, , \\
& \equiv & \exp \Big[ -i(\Omega_{m} - \frac{\Omega_{m}^{2}}{2\Omega}   )t \Big] D_{m}(t) \label{eq:W_uniform_Dicke}
\end{eqnarray}
where $\Omega_{m} \! = \! m\mathcal{A}/N$ is the Overhauser shift, $Z_{m} \! = \! \sum n_{j}$ is the partition function in the narrowed state, and $n_{j}$ is the number of $\beta$ states allowed for a given $j$ (see Refs.~\onlinecite{Arecchi_PRA72,Taylor_PRL03,Melikidze_PRB04,Ramon_PRB07}):
\beq
n_{j} = \binom{N}{N/2 - j} - \binom{N}{N/2-j-1} \,\, .
\eeq
Note that Eq.~(\ref{eq:W_uniform_Dicke}) in fact predicts the coherence to revive at Poincare time $t_{\text{P}} \! = \! \pi N^{2}\Omega/\mathcal{A}^2 \approx NT_{2,\text{short}} \gg T_{2,\text{short}} $, where we used $T_{2,\text{short}}$ from Eq.~(\ref{eq:T2}). Therefore this revival is inobservable in a realistic system, since other processes (such as spectral diffusion or even nuclear spin diffusion out of the dot) will alter the dynamics of the real system in the meantime and thus prevent the constructive interference of all the phases for different $j$ states.

\begin{figure}[t]
\centering
\includegraphics[width=0.9\linewidth]{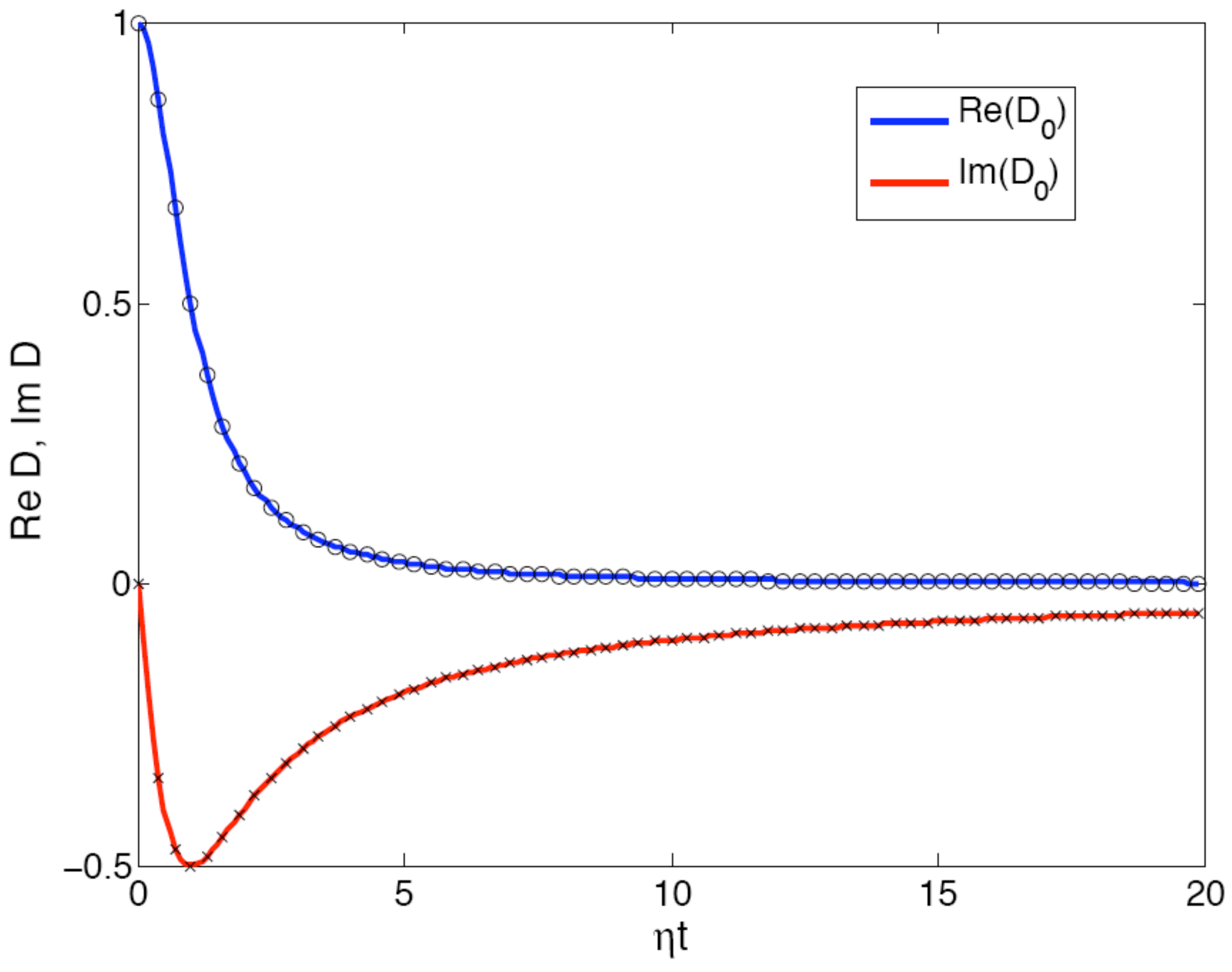}
  \caption{(Color online) FID decoherence in a system with uniform hf couplings $A_{i}$. The solid lines are the real and imaginary parts of $D_{m=0}$ from Eq.~(\ref{eq:W_uniform_Dicke}), which is the exact solution for decoherence due to the lowest-order hf-mediated interaction. The symbols are results of calculation within the ring diagram theory. } \label{fig:Dicke}
\end{figure}

On the other hand, the ring diagram solution in this case is given by
\begin{eqnarray}
W^{s}_{\text{rings}} (t)& = & e^{-i\Omega_{m}t} \frac{ \exp[- i \arctan \eta t) ]}{\sqrt{1+\eta^{2}t^{2}}} \,\, , \nonumber\\
& \equiv &  e^{-i\Omega_{m}t} D_{\text{rings}}(t) \label{eq:W_uniform_rings}
\end{eqnarray}
Note that for uniform hf couplings the above solution is valid at all times (as long as we only consider the $\tilde{H}^{(2)}$ interaction), so that there is no transition to exponential decay at long times.
Furthermore, within the ring diagram approach we do not discern different $m$ states (apart from the Overhauser shift $\Omega_{m}$): in the diagrammatic derivation we have assumed that all typical states (with $m \! < \! \sqrt{N}$) give the same $W(t)$ in the frame rotating with $\Omega + \Omega_{m}$ frequency.  In such a frame the expression from Eq.~(\ref{eq:W_uniform_Dicke}) still has the $\exp(-it\Omega^{2}_{m}/\Omega)$ phase factor which is absent in the ring diagram expression. Thus, the shifts of precession frequencies of the electron spin are different in the two calculations. Practically, this is not an issue since both can be identified with the experimentally observed frequency. The question is whether the non-trivial parts of the decoherence function, i.e.~$D_{m}$ and $D_{\text{rings}}$ agree with each other. Their comparison is shown in Fig.~\ref{fig:Dicke}, where we plot $D_{m=0}$ calculated numerically for $N \! = \! 2000$ nuclei, showing a very good agreement between the two calculations. This results supports our claim that the ring diagram resummation is a good solution for short times in realistic dots, since for $t\! \ll \! N/\mathcal{A}$ the exact shape of the wavefunction does not matter, and all the diagrams depend only on $N$ defined in Eq.~(\ref{eq:N}). 
We have also checked that as long as $m \! < \! \sqrt{N}$, the shape of $D_{m}$ remains practically unchanged at time-scale shown in Fig.~\ref{fig:Dicke}, confirming the result from Refs.~\onlinecite{Yao_PRB06,Liu_NJP07} that apart from different frequency shifts, all these narrowed states exhibit basically the same decoherence dynamics.

%%% GaAs FID figure
\begin{figure}[t]
\centering
\includegraphics[width=0.95\linewidth]{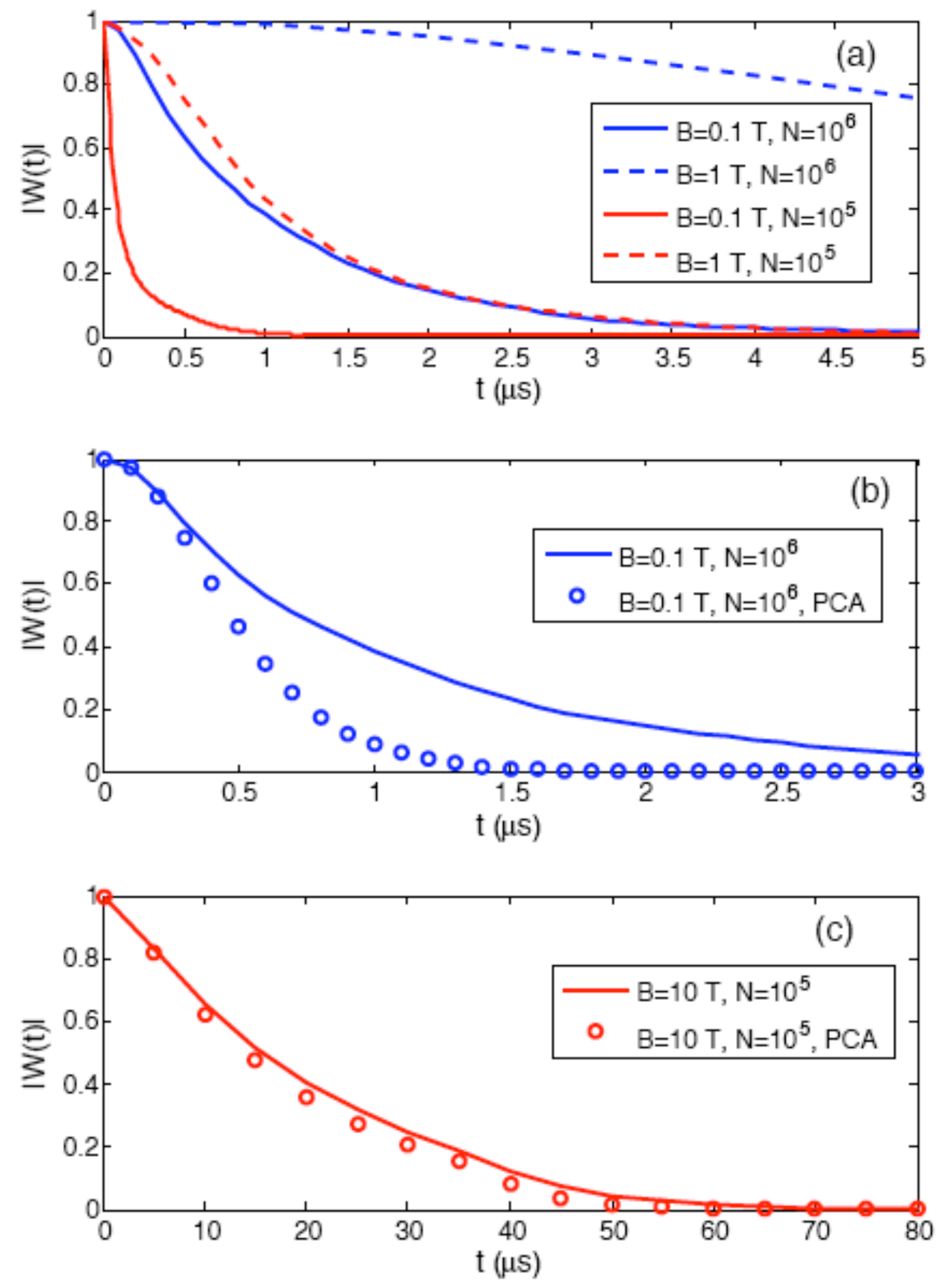}
  \caption{(Color online) Narrowed state $|W^{s}(t)|$ for FID calculated for two GaAs dot of different sizes and various magnetic fields. (a) Our theory calculated with coarse-grained $T$-matrix with $M_{A} \! = \! 10$.  (a) Red (blue) lines correspond to $N \! = \! 10^{5}\, (10^{6})$, while solid (dashed) lines correspond to $B \! =\! 0.1 \, (1)$ T ($g_{\text{eff}} \! = \! -0.44$ is used). Eq.~(\ref{eq:WFID_short}) corresponding to $M_{A} \! = \! 1$ gives practically the same results at this time-scale.
(b) Comparison of our result with all the ring diagrams resummed (solid line) with the pair correlation approximation (PCA) solution (circles) for $N \! = \! 10^{6}$ and $B \! = \! 0.1 $ T. The behavior of $|W(t)|$ after the initial decay (for $t \! > \! 1$ $\mu$s) is $\sim 1/t^{3}$ and $\sim \! \exp(-t^{2})$, respectively. (c) Analogous comparison for $N\! = \! 10^{5}$ and $B\! = \! 10$ T, when the decay occurs in the long time regime ($t \!\gg \! N/\mathcal{A}$), and the pair approximation is much closer to our result. 
  } \label{fig:GaAs_FID}
\end{figure}

\begin{figure}[t]
\centering
\includegraphics[width=0.95\linewidth]{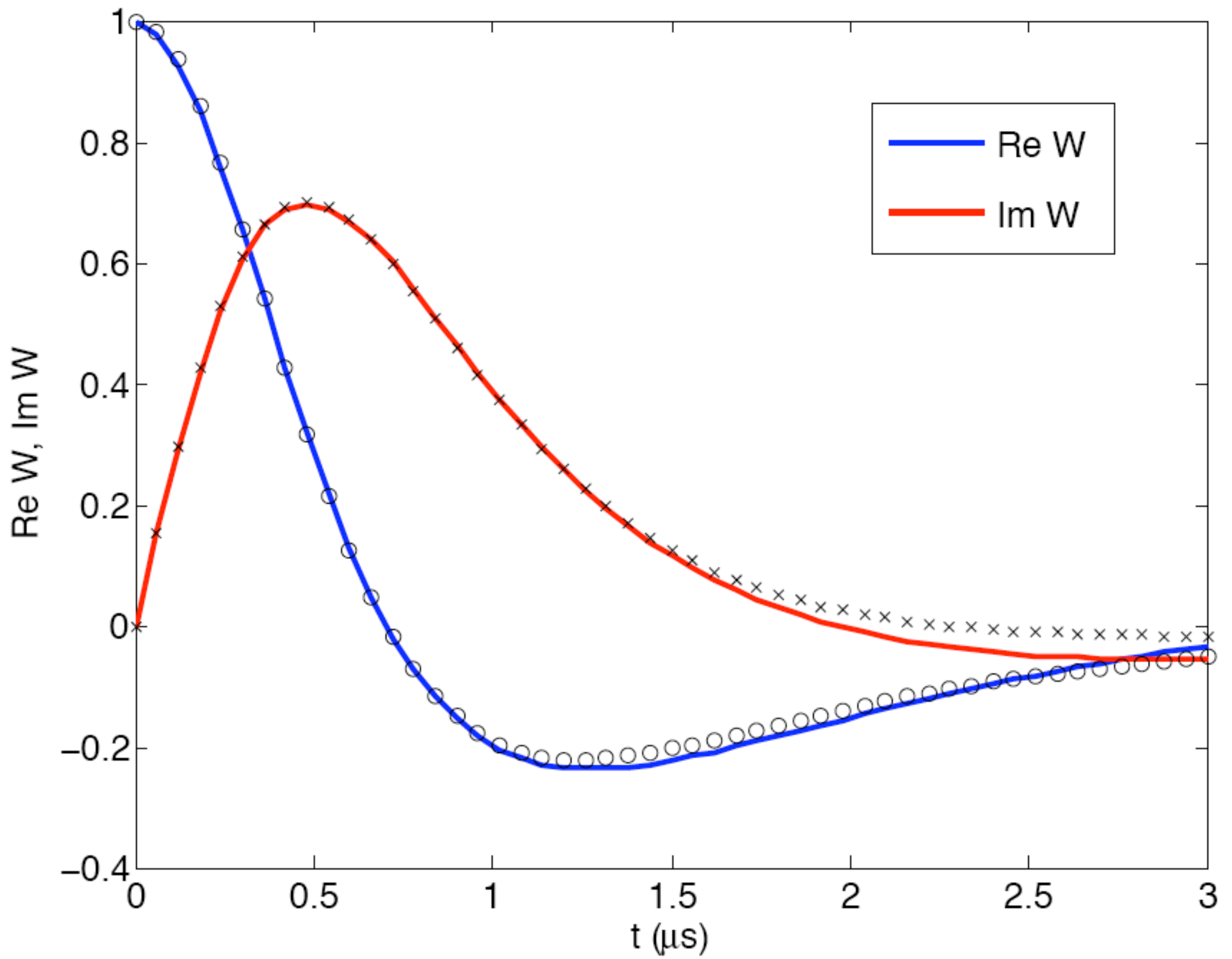}
  \caption{(Color online) Comparison between the numerical $T$-matrix calculation of $W^{s}_{\text{FID}}(t)$ and the analytical approximation from Eq.~(\ref{eq:WFID_short}). The solid lines are the real and imaginary part of $W^{s}(t)$, and the symbols are from Eq.~(\ref{eq:WFID_short}). The calculation is done for GaAs with $N\! = \! 10^{5}$ at $B\! = \! 1$ T. In the coarse-grained $\tilde{T}$-matrix calculation we have used $M_A \! =\! 30$. } \label{fig:phase}
\end{figure}

%%% InGaAs FID figure
\begin{figure}[t]
\centering
\includegraphics[width=0.9\linewidth]{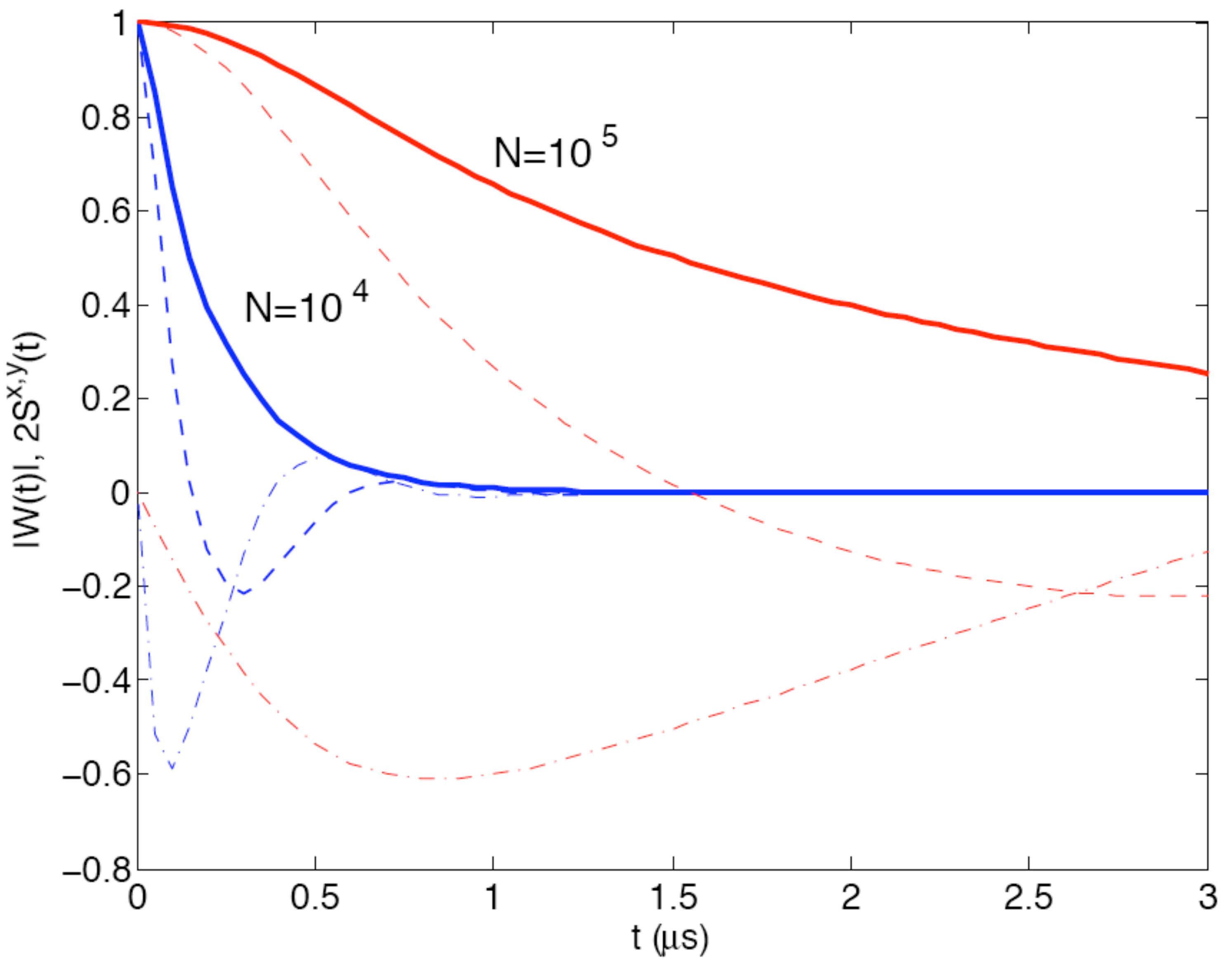}
  \caption{(Color online) FID decay in In$_{0.5}$Ga$_{0.5}$As for $N\! =\! 10^{4}$ (blue) and $10^{5}$ (red), at $B \! = \! 5$ T and with $g_{\text{eff}} \! = \! 0.5$.
  The solid lines are $|W(t)|$, while the dashed (dot-dashed) lines are the $S^{x}$ ($S^{y}$) components of the spin normalized to unity. We have assumed that at $t \! =\! 0$ the electron spin is initialized along the $x$ axis. 
  } \label{fig:InGaAs_FID}
\end{figure}

%%%%%%
%%%  FID RESULTS
%%%%%
\subsection{Results for GaAs and InGaAs}   \label{sec:FID_results}
In GaAs we have three nuclear species, $^{69}$Ga, $^{71}$Ga, and $^{75}$As, each having spin $J \! =\! 3/2$. The numbers of nuclei per unit cell are given by $n_{\alpha} \! = \! 0.604$, $0.396$, and $1$, respectively, and the hf energies $\mathcal{A}_{\alpha}$ are given in Table \ref{tab:energies}. In Fig.~\ref{fig:GaAs_FID}a we show $|W^{s}_{\text{FID}}(t)|$ for various $N$ and $B$ calculated using our theory, and in Fig.~\ref{fig:GaAs_FID}b-c we present their comparison with the PCA calculation.\cite{Yao_PRB06,Liu_NJP07} In Fig.~\ref{fig:GaAs_FID}b one can see that at low $B$, when the decay occurs at times shorter than $N/\mathcal{A}$,  there are visible differences between the fully resummed solution and the use of only two-spin ring diagram. At higher $B$ used in Fig.~\ref{fig:GaAs_FID}c, when the decay occurs at longer times, our calculation is in much closer agreement with the PCA results of Ref.~\onlinecite{Yao_PRB06}, since in the high $B$ and long-time regime the only difference between the two results is the presence of corrections to $T_{2,\text{long}}$ in our theory (which are quite small at $B\! = \! 10$ T in GaAs). The additional phase dynamics due to the ring diagram resummation is shown in Fig.~\ref{fig:phase}, where real and imaginary parts of $W^{s}_{\text{FID}}$ are plotted.

% InGaAs
The optical experiments on an ensemble of InGaAs dots have been interpreted\cite{Greilich_Science06} as measurement of narrowed-state FID (note, however, that the physics of the nuclei influenced by a train of optical pulses is richer and is still being investigated, see Refs.~\onlinecite{Greilich_Science07,Carter_PRL09}). The observed characteristic decay time was between $3$ $\mu$s for larger,\cite{Greilich_Science06} and $0.2$ $\mu$s for smaller dots.\cite{Carter_PRL09} In Fig.~\ref{fig:InGaAs_FID} we present our results for FID decay in In$_{x}$Ga$_{1-x}$As dot with $N \! = \! 10^{4}-10^{5}$ nuclei and indium concentration of $x\! = \! 0.5$. The parameters of the Hamiltonian are given in Table \ref{tab:energies}, the concentrations of $^{113}$In and $^{115}$In are $n_{\alpha} \! = 0.0428x$ and $0.9572x$, respectively, and both of these nuclei have spins $J \! =\! 9/2$. Note that for $N\! = \! 10^{4}$ practically the whole decay occurs in the long-time Markovian regime of exponential decay, but at the field of $5$ T we expect the long-time solution to be reliable.\cite{Coish_PRB08}
% ADDED:
The decay is practically completely dominated by interaction with In nuclei: due to their spin of $9/2$ they are much more efficient at decohereing the electron than $J\! = \! 3/2$ nuclei of Ga and As. 
We also remark that using the results from Ref.~\onlinecite{Witzel_PRB08} we have estimated the spectral diffusion decoherence time in such small InGaAs dots to be about $5$-$10$ $\mu$s, justifying our concentration on decoherence due to hf-mediated interactions which occurs on sub-microsecond time-scale.

%%%%%%%
%% SPIN ECHO - NONSECULAR TERMS
%%%%%%%
\section
{Spin Echo} \label{sec:SE}
It was first observed in numerical simulations\cite{Shenvi_scaling_PRB05} that at high $B$ practically all of the decoherence due to hf-mediated interaction was removed by the SE sequence. This result was explained in intuitive and transparent way using the pseudospin approach in the PCA.\cite{Yao_PRB06,Liu_NJP07} In these articles, the lowest-order $S^{z}$-conditioned interaction was treated together with the diagonal dipolar interaction ($ -2b_{ij} \hat{J}^{z}_{i}\hat{J}^{z}_{j} $ in Eq.~(\ref{eq:H_dip})), resulting in a near-perfect reversal of the nuclear spin dynamics by the SE sequence. The magnetic-field dependence of the small remaining decoherence was very weak\cite{Liu_NJP07} in the high $B$ regime considered there. It is crucial to note that in these papers the nuclear bath was assumed to be either homonuclear,\cite{Shenvi_scaling_PRB05} or the interactions between nuclei of different species were (justifiably) neglected\cite{Yao_PRB06,Liu_NJP07} at $B\! \approx \! 10$ T. 

The effect of the SE on decoherence due to the dipolar interactions  is much less dramatic, and it is now well established that in the very high magnetic field limit the SE decoherence is purely due to spectral diffusion.\cite{Witzel_PRB05,Witzel_PRB06,Yao_PRB06,Liu_NJP07,Witzel_AHF_PRB07} Here we consider much lower magnetic fields and concentrate only on the role of the hf-mediated interactions.

The following observation offers the key insight into which interaction channel is important in this case. If we remove all the nuclear intrabath interactions \emph{with exception of the $S^{z}$-conditioned terms} (i.e.~$\hat{V}_{1}$ in Eq.~(\ref{eq:HV1V2})), and treat these remaining terms in the secular approximation (allowing only for processes conserving the nuclear Zeeman energy), then there is no SE decoherence, i.e.~$W_{\text{SE}}(t)\! = \! 1$. Indeed, the nuclear interaction Hamiltonian in the secular approximation commutes with nuclear Zeeman energy $\HH_{Zn}$, and we have
\beq
e^{-i\tilde{H}_{\pm}t} = e^{-i( \HH_{Zn} \pm \HH_{A} \pm \hat{V}_{1})t } = e^{-iH_{Zn}t} e^{\pm i ( \HH_{A} + \hat{V}_{1})t } \,\, ,
\eeq
so that the product of evolution operators from Eq.~(\ref{eq:U_SE}) gives unity  in  Eq.~(\ref{eq:WU}). The same holds for any balanced dynamical decoupling sequence. 
Thus, there are three ways of getting finite SE decoherence from hf-mediated interactions. First is to reintroduce the dipolar interactions and consider the terms in the perturbation theory which mix them with the hf-mediated interaction (as in Refs.~\onlinecite{Yao_PRB06,Liu_NJP07}). Second is to consider the non-secular processes, and the the third way is to consider the $S^{z}$-independent hf-mediated interactions. 

Here we consider the two latter approaches for the following reasons. Mixing of the hf-mediated and dipolar interactions creates terms in the effective Hamiltonian which are partially local, and the number of terms in expansion of $W(t)$ due to these interactions does not scale as $N^{k}$ with the number $k$  of spins involved. Furthermore, a ring diagram (having all the nuclei distinct) with at least one $S^{z}$-conditioned interaction will still be equal to zero for SE. In order for the term in perturbation expansion to be non-zero, one has to consider the cases of repeating indices, e.g.~by doing the full diagrammatic calculation as in Ref.~\onlinecite{Saikin_PRB07}. The need to repeat the nuclear index in the summation brings us back to the ``term-counting'' argument. Finally, separating the dipolar and hf-mediated interactions allows for intuitive and transparent treatment: we can deal with the dipolar interaction using the well-tested cluster expansion theory,\cite{Witzel_PRB05,Witzel_PRB06,Witzel_PRB08} and treat the hf-mediated interactions separately using the ring diagram resummation which is appropriate for the case of long-range interactions.

% NONSECULAR LOWEST ORDER Sz-CONDITIONED
\subsection{T-matrix solution for Spin Echo}  \label{sec:SE_nonsec}
We repeat the calculations  from Sec.~\ref{sec:rings} using $f(t;\tau)$ for the SE is shown in Fig.~\ref{fig:contour}b.  
For the two-spin $S^{z}$-conditioned interaction from Eq.~(\ref{eq:H2_2s}) we get
\beq
T^{\text{SE}}_{kl} = B_{kl} \sqrt{a_{k}a_{l}}  \frac{8i\omega_{kl}e^{i\omega_{kl}t/2} \sin \frac{A_{kl}+\omega_{kl}}{4}t  \sin \frac{A_{kl}-\omega_{kl}}{4}t  }{A^{2}_{kl} - \omega^{2}_{kl} }  \,\, .      \label{eq:T_SE_Sz}
\eeq
As expected from the preceding discussion, we have $T_{kl} \! = \! 0$ when the two nuclei belong to the same species (i.e.~when $\omega_{kl} \! = \! 0$). At moderate $B$ fields (when $\omega_{kl} \! \gg \! A_{kl}$) and at short times we obtain for hetero-nuclear pairs 
\beq
T^{\text{SE}}_{kl}  \approx -B_{kl} \sqrt{a_{k}a_{l}}    \frac{8i}{\omega_{kl}} e^{i\omega_{kl}t/2}\sin^{2}\frac{\omega_{kl}t}{2} \,\, . \label{eq:T_SE_simple}
\eeq
The coarse-grained $T$-matrix from Eq.~(\ref{eq:T_coarse}) with $M_{J} \! =\! 1$ (expected to lead to accurate results at these short times) is then given by
\begin{eqnarray}
\tilde{T}_{\alpha\beta} & \equiv & (1-\delta_{\alpha\beta})  \sqrt{a_{\alpha}a_{\beta}} \sqrt{n_{\alpha}n_{\beta}} \frac{\mathcal{A}_{\alpha} \mathcal{A}_{\beta}}{N\Omega} \nonumber\\
& &\times  \frac{2i}{\omega_{\alpha\beta}} e^{i\omega_{\alpha\beta}t/2}\sin^{2}\frac{\omega_{\alpha\beta}t}{4} \, .  \,\,\,\,\,\,\,\,\,\,\,\label{eq:Tab}
\end{eqnarray}
With this we can write the ring diagram contribution  as 
\beq
R_{2k}(t) =  \sum_{i =1}^{N_{J}} \tilde{\lambda}^{2}_{i} \,\, ,  \label{eq:R_T_nonsec}
\eeq
where $\tilde{\lambda}_{i}$ are the eigenvalues of the $\tilde{T}$ matrix. As in Eq.~(\ref{eq:W_lambda}) we arrive then at the closed expression for the decoherence function due to the non-secular flip-flop processes:
\beq
W^{\text{SE}}_{\text{non-sec}}(t) = \prod_{i=1}^{N_{J}} \frac{1}{\sqrt{1+\tilde{\lambda}^{2}_{i} }} \,\, .
\eeq

\begin{figure}[t]
\centering
\includegraphics[width=0.9\linewidth]{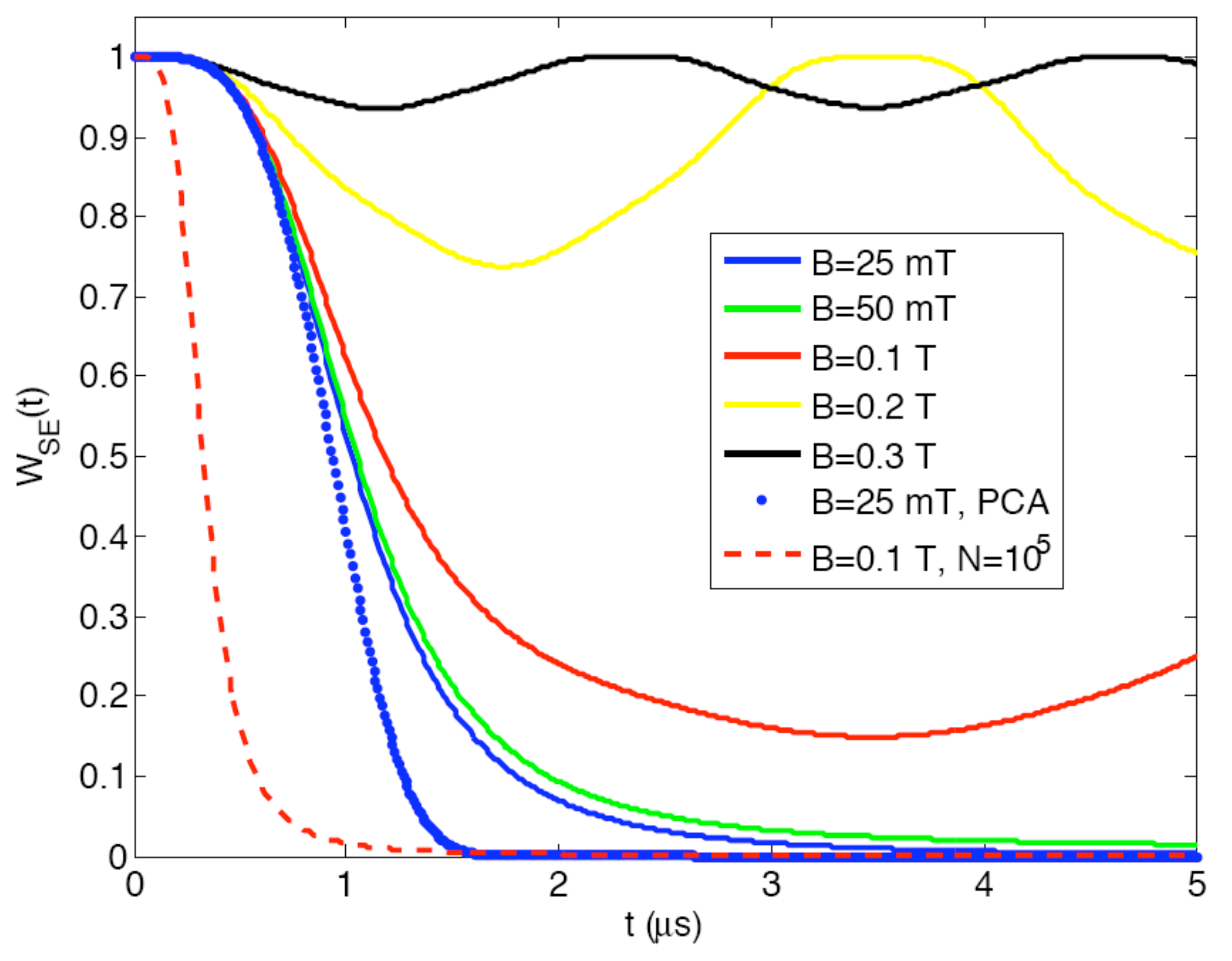}
  \caption{(Color online) Spin echo decoherence $W_{\text{SE}}(t)$ in GaAs at low $B$ fields. The parameters of GaAs are the same as used in FID calculations. Full $T$-matrix calculation was performed, but the results for $N \! =\! 10^{6}$ (solid lines) are indistinguishable from the approximation of Eq.~(\ref{eq:W_SE_R2}), while some differences between this analytical approximation and the numerics arise for the smaller dot with $N\! = \! 10^{5}$ (dashed line) at longer times (see Ref.~\onlinecite{Cywinski_PRL09}). The results for 25 mT and 50 mT show the saturation of decay at fields smaller than $B_{c}$ (see the text). The comparison of solid and dashed red lines (both for 0.1 T, but different $N$) illustrates how the low-$B$ decay scales with the dot size. The dots are the PCA result for 25 mT and $N\! = \! 10^{6}$, which should be compared with the solid blue line showing the result of calculation of all the ring diagrams. Similarly as in case of FID in Fig.~\ref{fig:GaAs_FID}b, PCA overestimates the decay beyond its initial stage.
  } \label{fig:SE}
\end{figure}

The case of $N_{J}\! = \! 3$ is experimentally relevant for GaAs quantum dots. There we obtain $\tilde{\lambda}_{i}(t) = 0, \,\, \pm \tilde{\lambda}(t)$
%\beq
%\tilde{\lambda}_{i}(t) = 0, \,\, \pm \tilde{\lambda}(t) \,\, ,
%\eeq 
which leads to 
\beq
W_{\text{SE}}(t) =   \frac{1}{1+\frac{1}{2}R_{2}(t) } \,\, ,  \label{eq:W_SE_R2}
\eeq
where we have identified $R_{2} \! = \! 2\tilde{\lambda}^2$, and $\tilde{\lambda}^{2}$ is given by
\beq
\tilde{\lambda}^{2}(t)  =  \left | \tilde{T}_{12}(t) \right |^{2} + \left | \tilde{T}_{13}(t) \right |^{2} + \left | \tilde{T}_{23}(t) \right |^{2} \,\, , \label{eq:lambda_T}
\eeq
and $\tilde{T}_{\alpha\beta}$ are given in Eq.~(\ref{eq:T_SE_simple}).
%\beq
%\left| \tilde{T}_{\alpha\beta} \right | ^{2} = \frac{4\mathcal{A}^{2}_{\alpha}\mathcal{A}^{2}_{\beta}}{N^{2}\Omega^{2}\omega^{2}_{\alpha\beta}} n_{\alpha}n_{\beta}a_{\alpha}a_{\beta} \sin^{4}\frac{\omega_{\alpha\beta}t}{4} \,\, . \label{eq:Tab_SE}
%\eeq

%%%%%%%
%%% Sz-independent pair flip-flop
%%%%%%%
Let us mention that analogously to the FID case, the higher-order two-spin interactions are giving negligible contributions compared to the one discussed above. 
For example, when we use the  $S^{z}$-independent term from Eq.~(\ref{eq:H3_2s}) as the interaction, we get the formula analogous to Eq.~(\ref{eq:T_SE_Sz}), but with the multiplicative factors of $\omega_{kl}$ in the numerator replaced by $A_{kl}$ and with $B_{kl}$ replaced by $C_{kl} \! = \! A_{k}A_{l}(A_{k}+A_{l})/16\Omega^2$.
Now both secular and non-secular terms are non-zero, but the secular terms are much larger for $\omega_{kl} \! \gg \! A_{kl}$, and the calculation of this contribution to decoherence parallels the one performed for FID in Sec.~\ref{sec:FID}. The result is that the corrections due to this interaction are completely negligible compared to the decoherence due to the lowest-order interaction for magnetic fields larger than 10 mT in a dot with $N \! =\! 10^{5}-10^{6}$ nuclei.

%%%%%%%%
%% SE RESULTS
%%%%%%%
\subsection{Spin Echo results and discussion}  \label{sec:SE_exp}
At moderate $B$ fields the SE decoherence is fully described by the two-spin ring diagram $R_{2}$ appearing as a result of resummation in the denominator of Eq.~(\ref{eq:W_SE_R2}). For GaAs $R_{2}(t)$ is given by a sum of three positive periodic functions, each having an amplitude $\sim \! B^{-4}$. Let us define the magnetic field $B_{c}$ for which these amplitudes become of the order of $1$. This $B_{c}$ corresponds to the electron Zeeman splitting of $\Omega_{c} \! \sim \!  \sqrt{r}\mathcal{A}/\sqrt{N}$, where $r \! = \! \Omega/\omega$, with $\omega$ being the typical difference of the nuclear Zeeman splittings, is of the order of $10^{3}$. 
Then, for  $B \! > \! B_{c}$ Eq.~(\ref{eq:W_SE_R2}) adds only a small oscillatory component to the total $W(t)$, and the actual SE decay is due to the spectral diffusion, which occurs on time-scale of more than $10$ $\mu$s for dots with $N \! \geq \! 10^{5}$, with the actual value depending on the shape of the dot.\cite{Witzel_PRB06,Witzel_PRB08} 

On the other hand, at $B\! < \! B_{c}$  (corresponding to $B\! < \! 250 (75)$ mT in GaAs with $|g_{\text{eff}}| \! =\! 0.44$ and $N\! = \! 10^{5} (10^{6})$), the decay due to the lowest-order hf-mediated interaction is significant. $R_{2}(t)$ is a sum of periodic functions, but their respective periods are incommensurate, and it is not expected for $R_{2}(t)$ to come back to $0$ (for $W(t)$ to come back to $1$) at a finite time $t$. Only partial ``rephasing'' occurs, and at  $B \! < \! B_{c}$ the minimal value of $R_{2}(t)$ becomes large and the time at which it is first achieved becomes long, so that the predicted SE decay is then from practical point of view irreversible, since spectral diffusion makes $W(t)$ decay to zero at a time-scale of $\geq \! 10$ $\mu$s anyway. The characteristic time of SE signal decay due to the hf-mediated interactions for $B\! < \! B_{c}$ is given by
\beq
T_{\text{SE}} \approx 3 \sqrt{r} \frac{\sqrt{N}}{\mathcal{A}} \approx 3 \sqrt{r} T^{\star}_{2} \,\, ,  \label{eq:TSE_lowB}
\eeq
which in GaAs translates to $T_{\text{SE}} \! \approx \! 100 T^{\star}_{2} $. 
These results are illustrated in Fig.~\ref{fig:SE} for GaAs dots with $N\! = \! 10^{6}$ and  $10^{5}$ nuclei. Eq.~(\ref{eq:TSE_lowB}) predicts the characteristic low-$B$ decay time of $\approx \! 0.5$ $\mu$s ($1.5$ $\mu$s) for $N\! = \! 10^{5}$ ($10^{6}$) in agreement with the results shown in Fig.~\ref{fig:SE}.

The calculated decay at low $B$ (when the signals cease to depend on the magnetic field) is in qualitative agreement with $T_{\text{SE}} \! \approx \! 0.3-0.4$ $\mu$s  measured in Ref.~\onlinecite{Koppens_PRL08} for $B\! \approx \! 50-70$ mT. The prediction of our theory is that at $B$ ten times larger the decay will be incomplete, having oscillations on a microsecond time-scale with a period proportional to $B$. 
Let us note that according to the theory presented here, the increase of $T_{\text{SE}}$ with increasing $B$ becomes visible only once $B$ becomes larger than $B_{c} \! = \! \Omega_{c}/g_{\text{eff}}\mu_{\text{B}}$, and it should be accompanied by an incomplete decay at longer times. At even higher $B$ the oscillations of the SE should become visible.
Let us note that it might be possible that the decay seen in Ref.~\onlinecite{Koppens_PRL08} was in fact incomplete (like the line line corresponding to $B \! =\! 0.1$ T and $N\! =  \! 10^{6}$ in Fig.~\ref{fig:SE}), since the measurements were done in a time window not much longer that the observed $T_{\text{SE}}$, and the exact value of signal (i.e.~current through the dot) corresponding to zero coherence could be uncertain.

\begin{figure}[t]
\centering
\includegraphics[width=0.9\linewidth]{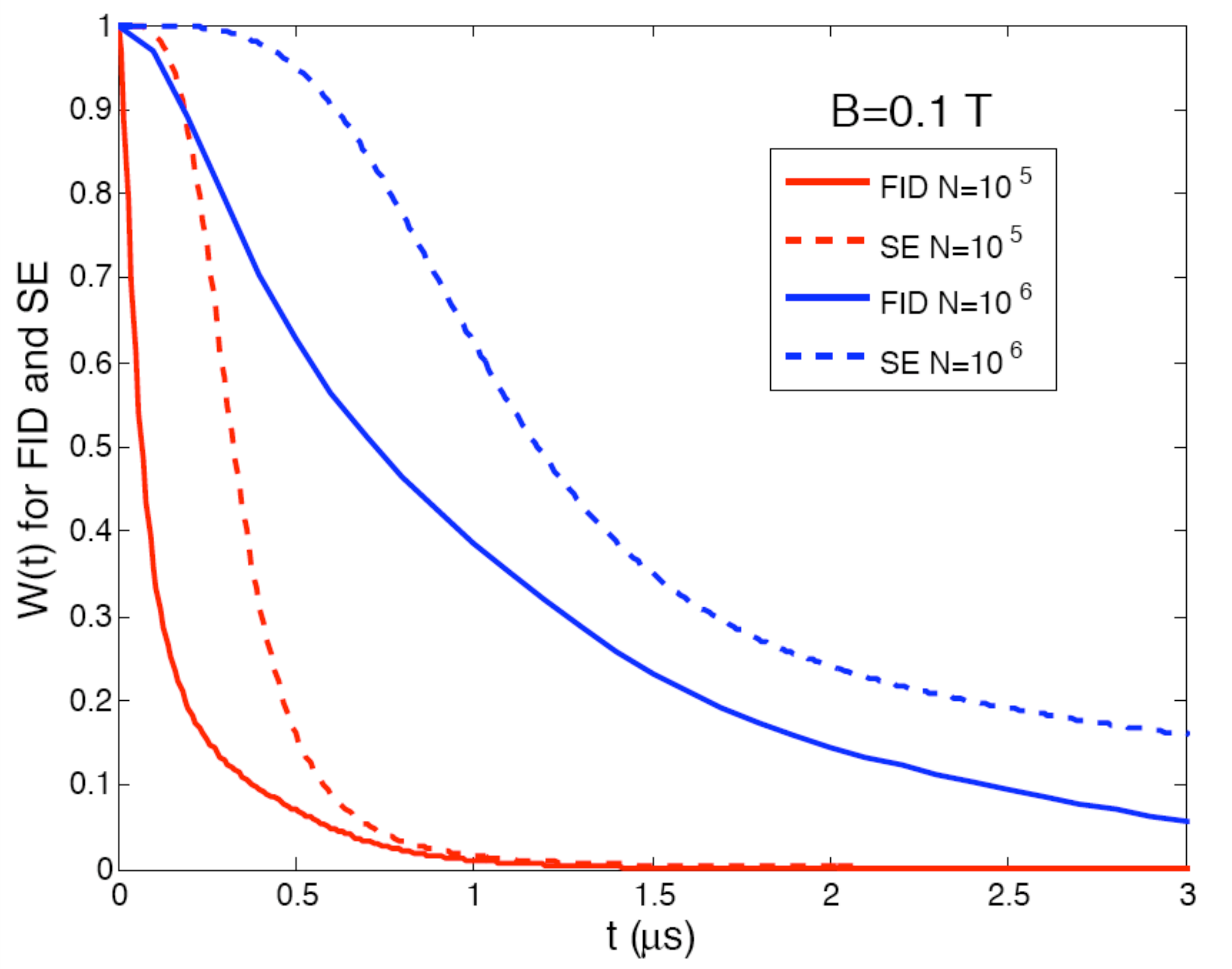}
  \caption{(Color online) Comparison between the single-spin FID decay and SE decoherence in GaAs quantum dot with $N \! =\! 10^{5}$ and $N\! =\! 10^{6}$ at $B \! =\! 0.1$ T. 
  } \label{fig:SE_FID}
\end{figure}

In Fig.~\ref{fig:SE_FID} we show a comparison of $W(t)$ for the single-spin FID and SE calculated for the same GaAs dot. As expected,\cite{Yao_PRB06} the coherence time is increased in the SE experiment compared to the single-spin FID. Note however that this is not entirely trivial, since the SE decay is dominated by processes of different character than the ones contributing to the FID dynamics.

%%%%%%%
%% DD
%%%%%%%
\section{Results for dynamical decoupling with multiple $\pi$ pulses} \label{sec:DD}
The calculation from the previous Section can be easily generalized to dynamical decoupling (DD) sequences\cite{Viola_JMO04,Khodjasteh_PRA07,Uhrig_PRL07} of many ideal $\pi$ pulses driving the electron spin. DD with various pulse sequences was considered theoretically for a spin bath both in a high $B$ and long time regime when the dipolar interactions play a large role,\cite{Yao_PRL07,Witzel_PRL07,Liu_NJP07,Witzel_CDD_PRB07,Lee_PRL08} and for very low $B$ fields when one has to use exact numerics.\cite{Zhang_Dobrovitski_PRB07,Zhang_Viola_PRB08} Here we focus on the heteronuclear system at low $B$  considered in Sec.~\ref{sec:SE}, and calculate the $W(t)$ for an experimentally realistic case  of applying a few pulses. As in Sec.~\ref{sec:SE}, we use a short-time approximation, so that only few first microseconds of coherence evolution are considered for GaAs with $N\! \approx \! 10^{6}$ nuclei, and we assume a moderate $B$ field at which $\omega_{\alpha\beta} \! \gg \! \mathcal{A}/N$. 
Only lowest-order hf-mediated interaction is considered, since it is the dominant source of the SE decay in this situation.  

At short times we have $W(t)$ given by Eqs.~(\ref{eq:W_SE_R2})-(\ref{eq:lambda_T}), only a different filter function $f(t;\tau)$ has to be used in calculation of $\tilde{T}_{\alpha\beta}$. It is convenient to rewrite $R_{2}(t)$ as
\beq
R_{2}= \sum_{\alpha} \sum_{\beta\neq\alpha} \mathcal{C}_{\alpha\beta}  \! \int_{0}^{t}d\tau_{1} \int_{0}^{t}d\tau_{2} f(t;\tau_{1}) e^{i\omega_{\alpha\beta}(\tau_{1}-\tau_{2})} f(t;\tau_{2}) \,\,   \label{eq:R2_C}
\eeq
with 
\beq
\mathcal{C}_{\alpha\beta} = a_{\alpha}a_{\beta}n_{\alpha}n_{\beta} \frac{\mathcal{A}^{2}_{\alpha}\mathcal{A}^{2}_{\beta}}{4\Omega^{2}N^{2}} \,\, .
\eeq
Let us introduce the ``spectral density'' function 
\beq
S(\omega) = 2\pi \sum_{\alpha} \sum_{\beta\neq\alpha} \mathcal{C}_{\alpha\beta} \delta(\omega + \omega_{\alpha\beta}) \,\, ,
\eeq
and also the filter function in the frequency domain defined by
\beq
F(\omega t) \equiv \frac{\omega^2}{2} \left | \tilde{f}(t;\omega) \right |^2 \,\, ,
\eeq
where $\tilde{f}(t;\omega)$ is the Fourier transform of $f(t;\tau)$ with respect to $\tau$. Then we can rewrite Eq.~(\ref{eq:R2_C}) as
\beq
R_{2} = 2 \int_{-\infty}^{\infty} \frac{d\omega}{2\pi} S(\omega) \frac{F(\omega t)}{\omega^{2}} \,\,  \label{eq:R2_S}
\eeq

The purpose of this formulation is the following. The pure dephasing decoherence due to $\hat{S}^{z}\beta(t)$ coupling to classical Gaussian noise $\beta(t)$ having spectral density $S(\omega)$ is given by $W(t) \! =\! \exp[-R_{2}(t)/2]$ with $R_{2}$ given by Eq.~(\ref{eq:R2_S}). For derivation see e.g.~Refs.~\onlinecite{deSousa_review,Cywinski_PRB08}. Thus, at very short times, when $R_{2} \! \ll \! 1$, the decoherence due to hf-mediated interactions between different nuclear species in regime of interest can be mapped onto decoherence due to classical Gaussian noise. It is interesting to note that a similar formal equivalence between initial decoherence of a quantum model under pulses sequences and classical noise problem has been derived for a very different bath Hamiltonian in Ref.~\onlinecite{Lutchyn_PRB08}.
Furthermore, for the case at hand we have Eq.~(\ref{eq:W_SE_R2}), which allows us to use Eq.~(\ref{eq:R2_S}) to describe the decoherence at longer times, at which the analogy to the classical Gaussian noise does not hold.

\begin{figure}[t]
\centering
\includegraphics[width=0.9\linewidth]{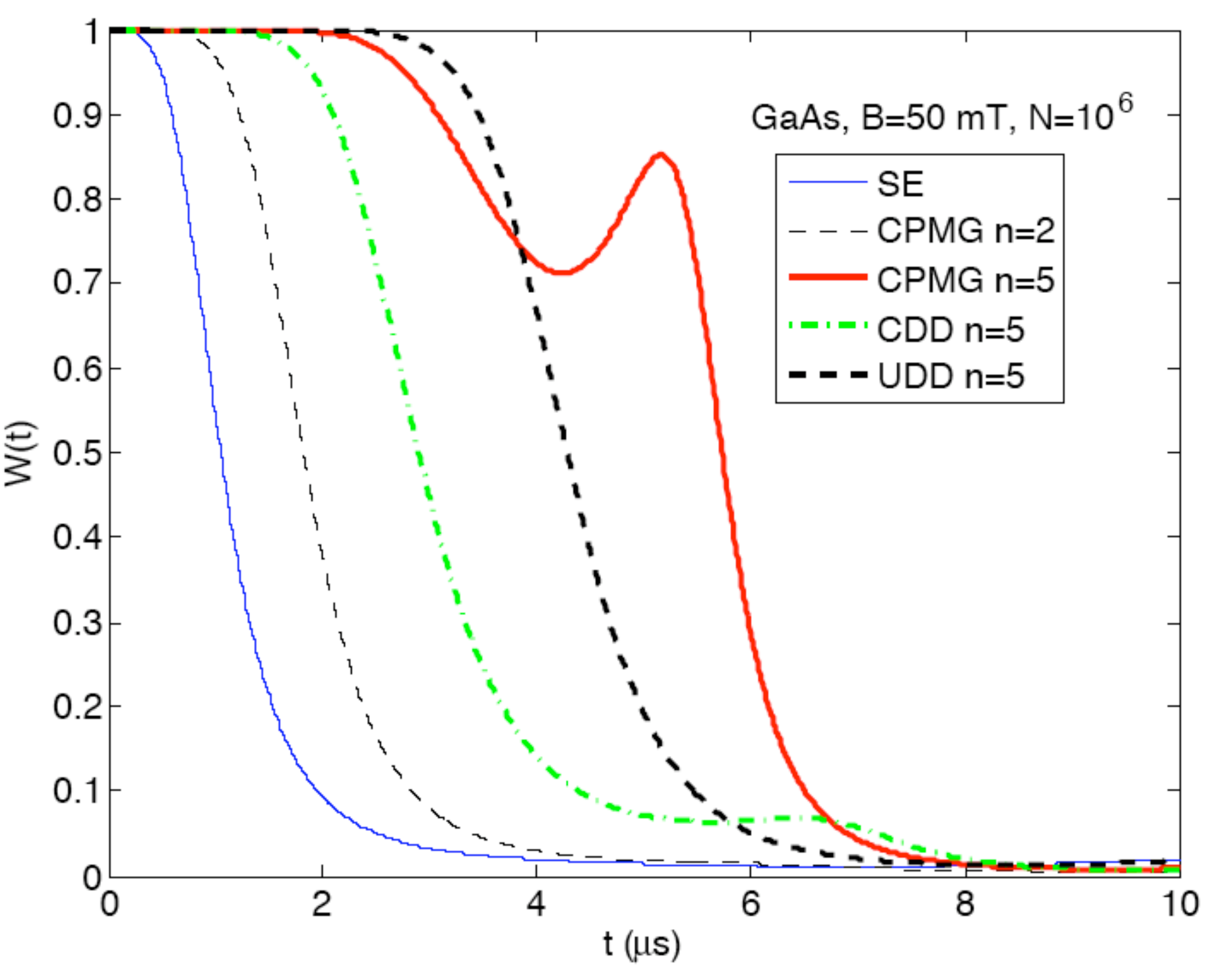}
  \caption{(Color online) Decoherence under various dynamical decoupling pulse sequences in  GaAs quantum dot with $N \! =\! 10^{6}$ at $B\! =\! 0.25$ T. The results for SE, 2-pulse CPMG sequence, and 5-pulse CPMG, CDD, and UDD sequences are shown.
  } \label{fig:DD}
\end{figure}

The frequency domain filter functions $F(\omega t)$ were calculated in Ref.~\onlinecite{Cywinski_PRB08} for different pulse sequences aimed at preventing pure dephasing, namely the CPMG sequence,\cite{Haeberlen} the concatenations of spin echo,\cite{Khodjasteh_PRA07,Yao_PRL07,Witzel_CDD_PRB07} the periodic sequence of $\pi$ pulses,\cite{Viola_JMO04} and the Uhrig's sequence.\cite{Uhrig_PRL07,Lee_PRL08,Yang_PRL08} For example, for SE we have $F(z) \! = \! 8\sin^{4}[z/4]$, which leads us back to the results obtained in Sec.~\ref{sec:SE}. The two pulses CPMG, concatenated spin echo (CDD), and Uhrig's sequence (UDD) all correspond to $t/4-\pi-t/2-\pi-t/4$ sequence (see Fig.~\ref{fig:contour}b), with the filter
\beq
F(z) = 128\cos^{2}\frac{z}{8}\sin^{6}\frac{z}{8} \,\, ,
\eeq
and $F(z)$ functions for sequences with more pulses are given in Ref.~\onlinecite{Cywinski_PRB08}.

The common feature of all the DD sequences is that their $F(\omega t)$ functions are filtering out large part of $S(\omega)$ in Eq.~(\ref{eq:R2_S}) at frequencies smaller than $\sim \! n/t$, where $n$ is the number of pulses.  
In fact, the Uhrig's sequence (UDD) is most efficient at this task when the noise spectrum has an upper frequency cutoff,\cite{Lee_PRL08,Cywinski_PRB08,Uhrig_NJP08}  which is exactly the case here (the cutoff being the maximal $|\omega_{\alpha\beta}|$). 

In Figure \ref{fig:DD} we present the results obtained for $W(t)$ in GaAs for DD sequences with up to $5$ pulses. As expected, the UDD sequence makes $W(t)$ very close to $1$ for the longest time. The characteristic features visible for times at which $(W(t)$ is smaller, i.e.~the peak in signal for five pulse CPMG sequence and a plateau for the CDD sequence, originate from the discrete nature of the noise spectrum. The peak in $W(t)$ for $n\!=\! 5$ CPMG comes from the $\omega_{\alpha\beta}t$ factor (with $\omega_{\alpha\beta}$ frequency for $^{71}$Ga and $^{75}$As nuclei) matching a minimum of the $F(\omega t)$ filter at $\omega t \! \approx  3\pi$. The plateau in the CDD signal has similar origin. In fact, such features have been observed in a recent study of dynamical decoupling from classical noise in ion trap qubits,\cite{Biercuk_Nature09} where a plateau in $W(t)$ was related to the presence of sharp peak in $S(\omega)$.

%%%%%%%
%% HIGHER ORDER CORRECTIONS
%%%%%%%
\section{Corrections due to higher order terms in the effective Hamiltonian} \label{sec:higher}
In the preceding sections we have presented calculations of decoherence due to hf-mediated interactions involving pairs of spins. Among such interactions in the effective Hamiltonian $\tilde{H}$, the most important one was the lowest-order $S^{z}$-conditioned one from Eq.~(\ref{eq:H2_2s}), and the contribution of higher-order two-spin interactions was shown to be very small using the ring diagram approach. What remains to be shown is that the influence of higher-order multi-spin interactions (such as the three-spin one in Eq.~(\ref{eq:H3_3s})) is small at the time-scale of coherence decay.

First, let us make a general remark about the convergence of the operator series $\tilde{H}^{(n)}$.
If one applies a commonly used operator norm (see e.g.~Ref.~\onlinecite{Khodjasteh_PRA07})
\beq
|| \hat{A} ||_{2} \equiv \max_{\bra{\Phi}\Phi\rangle=1} | \bra{\Phi} \hat{A} \ket{\Phi} | \,\, ,\label{eq:norm}
\eeq
to the leading $\tilde{H}^{(n)}$ terms (i.e.~terms containing $n$ spin operators), one gets an estimate 
\beq
|| \tilde{H}^{(n)} ||_{2} \approx \mathcal{A} \left( \frac{\mathcal{A}}{\Omega} \right)^{n-1} \,\, ,
\eeq
so that it seems that the expansion is convergent only when $\mathcal{A}/\Omega \! < \! 1$. 
However, the state $\ket{\Phi}$ which maximizes Eq.~(\ref{eq:norm}) is a highly entangled state of nuclear spins. Here we are interested in evolution of nuclear spin system averaged using an uncorrelated thermal ensemble, and such highly entangled states do not appear in the calculations. Thus, it is plausible that the condition for convergence (in the sense of giving us well defined results for $W(t)$) of the expansion of $\tilde{H}$, is much less restrictive than $\mathcal{A}/\Omega \! < \! 1$.

The full treatment of decoherence due to leading order terms in $\tilde{H}$ would require a derivation of a diagrammatic technique involving multi-spin interactions. Instead of pursuing this path, we choose to give a more limited, but practically sufficient argument. We calculate the lowest order contribution to $W(t)$ due to a multi-spin interaction, e.g.~for three-spin (3s) interaction from Eq.~(\ref{eq:H3_3s}) we obtain 
\beq
W^{(2)}_{3s} = \sum_{k} \frac{A^{2}_{k}}{\Omega^{2}} \frac{a^{2}_{k}}{2} W^{(2)}_{2s} \approx \delta^{2}W^{(2)}_{2s} \,\, .
\eeq
and although the analogous expressions for $W^{(2)}$ due to interactions involving more than three spins get more complicated, they are suppressed by higher powers of $\delta$ compared to $W^{(2)}_{2s}$.
In fact, on the time-scale of $t \! \ll \! N/\mathcal{A}$, while $W^{(2)}_{2s}$ can be larger than $1$ (thus forcing us to resum the higher-order terms), for $\delta\! \ll \! 1$ the lowest-order contributions from multi-spin interactions are still negligible, and it is not necessary to go to the higher orders of perturbation theory to conclude that these interactions are irrelevant \emph{on this time-scale}.
Note that the key results of this paper were obtained for magnetic fields at which the coherence was decaying significantly in the  short time regime.
Thus, we arrive at conclusion  that as long as $\delta \! \ll \! 1$, the contribution of multi-spin interactions to the full $W(t)$ is very small \emph{on the time-scale on which the coherence decay due to the two-spin interaction occurs at low fields}. The question of convergence of our approach at longer times, relevant for smaller $N$ or larger $B$ (but not so large as to give $\mathcal{A}/\Omega \! < \! 1$, when the convergence of the effective Hamiltonian expansion becomes more obvious), is left for future investigation.

%%%%%%%
%% CONCLUSION
%%%%%%%
\section{Conclusion}
We have presented a detailed description of the theory of pure dephasing decoherence of an electron spin coupled to  a nuclear bath by hyperfine interaction. At finite magnetic field $B$ one can perform a canonical transformation which eliminates the direct electron-nuclear spin-flip terms in hf interaction in favor of an effective pure dephasing Hamiltonian containing hf-mediated long-range interactions between the nuclei. 
We have argued that when the bath is thermal (uncorrelated and unpolarized), one can use the lowest order hf-mediated interaction to calculate the spin decoherence as long as the electron spin splitting $\Omega$ is larger than the rms of the Overhauser field distribution $\sim \mathcal{A}/\sqrt{N}$ (this correspond to $B \! \gg \! 3$ mT in GaAs dots with $10^{6}$ nuclei). 
The solution is possible due to long-range nature of the hf-mediated interactions, which couple with comparable strength all the $N$ nuclei within the bulk of the electron's wave-function ($N$ being larger than $10^{4}$ in III-V based quantum dots). In such a case we can identify the leading (in terms of $1/N$ expansion) diagrams in the perturbative series for the decoherence function, and then resum all these \emph{ring diagrams}. At short times (when $t \! \ll N/\mathcal{A}$, where $\mathcal{A}$ is the hf coupling energy) we can perform this resummation analytically and obtain closed formulas for decoherence in Free Induction Decay (FID), Spin Echo (SE), or any other dynamical decoupling sequence of pulses driving the electron spin. At longer times, the solution can be obtained numerically at very low computational cost.

Compared to previous work on decoherence due to hf-mediated interactions,\cite{Yao_PRB06,Liu_NJP07,Saikin_PRB07,Coish_PRB08} we have accounted for all the leading terms in the diagrammatic expansion of decoherence time-evolution function. This allows us to firmly establish the regimes in which the pair correlation approximation\cite{Yao_PRB06,Liu_NJP07} is quantitatively applicable, and also to make comparisons with results obtained using a very different Generalized Master Equation approach.\cite{Coish_PRB08} 
The resummation of all the ring diagrams leads to modified results for single-spin FID (i.e.~FID measured in a narrowed nuclear state with well-defined polarization) decay at low magnetic fields compared to the pair-correlation approximation (PCA) results,\cite{Yao_PRB06,Liu_NJP07} while at high fields we recover the PCA solution given in these papers. We also found nontrivial phase dynamics of the spin coherence in the FID experiment. The ring-diagram approach to FID was tested by showing that it agrees with an exact analytical solution in the case of uniform hf couplings. 

For SE decoherence, apart from including all the ring diagrams in the calculation, we have identified the most important mechanism of SE decay in III-V quantum dots at low magnetic fields: the hf-mediated interaction between spins of different nuclear species. While at high $B$ this interaction channel leads only to a small modulation of the SE signal, at low $B$ used in recent SE experiments\cite{Koppens_PRL08} on GaAs dots this leads to coherence decay on a sub-microsecond time-scale. This result is in qualitative agreement with the measurement from Ref.~\onlinecite{Koppens_PRL08}. We have also given predictions for the effect of dynamical decoupling pulse sequences on decoherence in the regime in which the SE has been measured. 

We believe that this work extends the  analytical theory of spin decoherence due to interaction with the nuclei down to magnetic fields much smaller than previously assumed, and leaves only the highly nontrivial zero field limit to exact numerical approaches. Using the theory presented here one can make qualitative and quantitative predictions, particularly in  the experimentally relevant regimes of low magnetic fields and relatively short time scales.

\begin{acknowledgments}
We thank  W.A.~Coish,  V.V.~Dobrovitski, S.E.~Economou, Al.L. Efros, D.~Gammon, A.~Greilich, X.~Hu, R.-B. Liu, S.K.~Saikin, and A.~Shabaev for discussions and useful remarks. This work is supported by LPS-NSA.  
\end{acknowledgments}

\appendix
%%%%%%
% APPENDIX: CANONICAL TRANSFORMATION
%%%%%%
\section{Canonical transformation leading to the effective pure-dephasing Hamiltonian} \label{app:canonical}
Following Ref.~\onlinecite{Winkler} we divide the Hilbert space of our problem into two parts, A and B, spanned by eigenstates of $\HH_{\text{Z}}$, and we assume that for all eigenstates $m\in A$ and $l\in B$ we have $E^{Z}_{m}-E^{Z}_{l}$$\approx$$\Omega$, with $\Omega$ being much larger than the spread of energies within $A$ and $B$ and the energy scale of perturbations which we will consider. 

The full Hamiltonian is then written as
\beq
\HH = \HH_{0} + \HH_{1} + \HH_{2} = \HH_{0} + \HH{}'
\eeq
where $\HH_{0} \! = \! \Omega \hat{S}^{z}$, $\HH_{1}$ has only matrix elements within the A and B subsets of states (i.e. the diagonal hf coupling $A_{i}J^{z}_{i}\hat{S}^{z}$, nuclear Zeeman, and dipolar interactions are contained in $\HH_{1}$), and $\HH_{2}$  has only matrix elements between the states from blocks A and B (and actually $\HH_{2} \! = \! \Vsf$, the spin-flip part of the hf interaction).
We want to devise a transformation which lead to a new Hamiltonian $\tilde{H}$ which has no matrix elements between the states from A and B subspaces. We write this Hamiltonian as
\beq
\tilde{H} = e^{-\mathcal{\hat{S}}} \HH e^{\mathcal{\hat{S}}} = \sum_{n=0}^{\infty} \frac{1}{n!} [\HH,\mathcal{\hat{S}}]^{(n)}\,\, ,  \label{eq:H_nested}
\eeq
where $\mathcal{\hat{S}}$ is an anti-Hermitian operator, and we have defined the nested commutator:
\beq
[\HH,\mathcal{\hat{S}}]^{(n)} \equiv [...[[\HH,\mathcal{\hat{S}}],\mathcal{\hat{S}}], .....\mathcal{\hat{S}}] \,\, ,
\eeq
with $\mathcal{\hat{S}}$ appearing $n$ times.

%ADDED:
We refer to the operators having nonzero matrix elements only within the A and B subspaces as block diagonal (BD), and to those having nonzero matrix elements only between the states from A and B subspaces as block off-diagonal (BOD). A product of two BD or two BOD operators is a BD operator, while a product of a BD and a BOD operator is a BOD operator. In order to obtain a closed hierarchy of equations for $\mathcal{\hat{S}}$ we need to assume that $\mathcal{\hat{S}}$ is a BOD operator. 
Then the condition for the BOD part of $\tilde{H}$ to vanish is
\beq
\sum_{n=0}^{\infty} \frac{1}{(2n+1)!} [\HH_{0}+\HH_{1},\mathcal{\hat{S}}]^{(2n+1)} + \sum_{n=0}^{\infty} \frac{1}{(2n)!} [\HH_{2},\mathcal{\hat{S}}]^{(2n)} = 0 \,\, , \label{eq:off}
\eeq
and the non-zero (BD) part of $\tilde{H}$ is given by
\beq
\tilde{H} = \sum_{n=0}^{\infty} \frac{1}{(2n)!} [\HH_{0}+\HH_{1},\mathcal{\hat{S}}]^{(2n)} + \sum_{n=0}^{\infty} \frac{1}{(2n+1)!} [\HH_{2},\mathcal{\hat{S}}]^{(2n+1)} \,\, , \label{eq:diag}
\eeq
Now we expand $\mathcal{\hat{S}}$ into operators of increasing order in $\HH{}'$:
\beq
\mathcal{\hat{S}} = \mathcal{\hat{S}}^{(1)} + \mathcal{\hat{S}}^{(2)} + ...
\eeq
and by plugging this expansion into Eq.~(\ref{eq:off}) we obtain an infinite set of equations:
\begin{align}
[\HH_{0},\mathcal{\hat{S}}^{(1)}] & =  -\HH_{2} \,\, , \\
[\HH_{0},\mathcal{\hat{S}}^{(2)}] & =  -[\HH_{1},\mathcal{\hat{S}}^{(1)}] \,\, , \\
[\HH_{0},\mathcal{\hat{S}}^{(3)}] & =  -[\HH_{1},\mathcal{\hat{S}}^{(2)}] - \frac{1}{3}[[\HH_{2},\mathcal{\hat{S}}^{(1)}],\mathcal{\hat{S}}^{(1)}] \,\, ,  \text{etc...}
\end{align}
Now we define the operator $\hat{\Delta}$:
\beq
\hat{\Delta} = \hat{P}_{A} - \hat{P}_{B} \,\, ,
\eeq
in which $\hat{P}_{A(B)}$ is the projection on the subspace A(B). 
With the energy separation between the subspaces:
\beq
\Omega \approx E^{0}_{A} - E^{0}_{B} \,\, ,
\eeq
where $E^{0}_{A,B}$ are the typical zeroth-order energies (i.e.~the diagonal elements of $\HH_{1}$), we arrive at
\beq
[\HH_{0},\hat{X}] \approx -\Omega \hat{X} \hat{\Delta} \,\, .  \label{eq:app_Delta}
\eeq
for $\hat{X}$ being a purely block-off-diagonal operator (such as $\mathcal{\hat{S}}$). This approximation corresponds to neglecting terms of the order of $\mathcal{A}/\sqrt{N}$ and $\omega_{i}$ compared to $\Omega$. 
% ADDED: !!!
This will affect the values of coupling constants in certain terms in $\tilde{H}^{(n)}$ with $n\! > \! 2$, but it will not alter the structure of multi-spin nuclear interactions in $\tilde{H}$ and the scaling of their coupling energies with $\mathcal{A}$ and $\Omega$. 
The formulas for the second-order effective Hamiltonian derived without making this approximation are given in e.g.~Refs.~\onlinecite{Shenvi_scaling_PRB05,Coish_PRB08}. 

Using the following properties of $\hat{\Delta}$:
\beq
\hat{\Delta}^{2}  = 1 \,\,\, , \,\,\,\,\, [\HH_{1},\hat{\Delta}] = 0 \,\,\, , \,\,\,\,\, \{\HH_{2},\hat{\Delta}\} = 0 \,\, ,
\eeq
we obtain the explicit formulas for $S^{(n)}$:
\begin{align}
\mathcal{\hat{S}}^{(1)} &= \frac{1}{\Omega} \HH_{2} \hat{\Delta} \,\, , \\
\mathcal{\hat{S}}^{(2)} &= \frac{1}{\Omega^{2}} [\HH_{1},\HH_{2}]  \,\, , \\
\mathcal{\hat{S}}^{(3)} &= \frac{1}{\Omega^{3}} [\HH_{1},[\HH_{1},\HH_{2}]]\hat{\Delta} + \frac{1}{3\Omega^{3}}[[\HH_{2},\HH_{2}\hat{\Delta}],\HH_{2}\hat{\Delta}]\hat{\Delta}
\end{align}
Using these, from Eq.~(\ref{eq:diag}) we can $\tilde{H}$ up to the fourth order in $\HH{}'$:
\begin{eqnarray}
\tilde{H}^{(1)} &=& \HH_{1} \,\, ,\\
\tilde{H}^{(2)} &= &\frac{1}{2}[\HH_{2},\mathcal{\hat{S}}^{(1)}] = \frac{1}{2\Omega}[\HH_{2},\HH_{2}\hat{\Delta}] = \frac{1}{\Omega} \HH_{2}^{2}\hat{\Delta}, ~~~~ \label{eq:H2S1} \\
\tilde{H}^{(3)} &= & \frac{1}{2}[\HH_{2},\mathcal{\hat{S}}^{(2)}] = \frac{1}{2\Omega^{2}} [\HH_{2},[\HH_{1},\HH_{2}]] \,\, ,\\
\tilde{H}^{(4)} &= & \frac{1}{2}[\HH_{2},\mathcal{\hat{S}}^{(3)}] - \frac{1}{24}[[[\HH_{2},\mathcal{\hat{S}}^{(1)}],\mathcal{\hat{S}}^{(1)}],\mathcal{\hat{S}}^{(1)}]  \nonumber\\
& = &\frac{1}{\Omega^{3}} \Big( \frac{1}{2}\HH_{1}^{2}\HH_{2}^{2} +  \frac{1}{2}\HH_{2}^{2}\HH_{1}^{2} -\HH_{2}\HH_{1}\HH_{2}\HH_{1} \nonumber\\
& & - \HH_{1}\HH_{2}\HH_{1}\HH_{2} + \HH_{2}\HH_{1}^{2}\HH_{2} -\HH_{2}^{4} \Big)  \hat{\Delta} \,\, .
\end{eqnarray}
In higher orders the formulas get even more complicated. However, for $n$ odd we always have expressions involving products of even number of $\HH_{2}$ and odd number of $\HH_{1}$, whereas for $n$ even we get $\tilde{H}^{(n)}$ being a sum of products of even numbers of $\HH_{1}$ and $\HH_{2}$ operators, multiplied then by $\hat{\Delta}$.

The full expression for $\tilde{H}^{(2)}$ is given in Sec.~\ref{sec:Heff}, where also certain terms from $\tilde{H}^{(3)}$ were shown. Here we give all the terms in $\tilde{H}^{(3)}$ which arise when we neglect the dipolar interactions in $\HH_{1}$ (i.e.~we keep only the Zeeman and diagonal hf term in $\HH_{1}$):
\begin{widetext}
\begin{eqnarray}
\tilde{H}^{(3)}_{\text{hf}} & = & \hat{S}^{z} \sum_{i} \frac{A_{i}^{3}}{4\Omega^2} \hat{J}^{z}_{i} 
+ \sum_{i,j} \frac{A_{i}A_{j}(A_{i}+A_{j})}{8\Omega^2} \hat{J}^{z}_{i}\hat{J}^{z}_{j} 
- \sum_{i}\frac{A_{i}^{3}}{8\Omega^2} \left( \hat{J}^{2}_{i} - (\hat{J}^{z}_{i})^2 \right) 
- \sum_{i\neq j}  \frac{A_{i}A_{j}(A_{i}+A_{j})}{16\Omega^2} \hat{J}^{+}_{i}\hat{J}^{-}_{j} \nonumber\\
& & - \hat{S}^{z} \sum_{i\neq j} \frac{A^{2}_{i}A_{j}}{2\Omega^{2}}  \left( \hat{J}^{2}_{i} - (\hat{J}^{z}_{i})^2 \right) \hat{J}^{z}_{j} - 
 \hat{S}^{z} \sum_{i\neq j} \frac{A^{2}_{i}A_{j}}{2\Omega^{2}}  \hat{J}^{z}_{i} (  \hat{J}^{+}_{i}\hat{J}^{-}_{j} +  \hat{J}^{+}_{j}\hat{J}^{-}_{i} ) 
 - \hat{S}^{z} \sum_{i\neq j} \frac{A^{2}_{i}A_{j}}{4\Omega^{2}} (  \hat{J}^{-}_{i}\hat{J}^{+}_{j} -  \hat{J}^{+}_{i}\hat{J}^{-}_{j} ) \nonumber\\
 & & -\hat{S}^{z} \sum_{i\neq j \neq k} \frac{A_{i}A_{j}A_{k}}{2\Omega^{2}}   \hat{J}^{+}_{i}\hat{J}^{-}_{j} \hat{J}^{z}_{k} \,\, .  \label{eq:H3_nonlocal}
\end{eqnarray}
\end{widetext}
The last term above is the three-spin interaction from Eq.~(\ref{eq:H3_2s}). Terms such as this one, which contain $n$ nuclei in $n$-th order of expansion of $\tilde{H}$, are potentially the most ``dangerous'' ones when one considers the convergence of the expansion. For example, in the the next order we have such a ``maximally nonlocal'' term given by
\beq
\tilde{H}^{(4)}_{4s}  =  -\hat{S}^{z} \!\!\!\! \sum_{i\neq j\neq k \neq l}  \!  \frac{A_{i}A_{j}A_{k}A_{l}}{16\Omega^{3}} \hat{J}^{+}_{i}\hat{J}^{-}_{j}\hat{J}^{+}_{k}\hat{J}^{-}_{l} \,\, . \label{eq:H4_4s}
\eeq

%%%%
%%% REPLICA PROOF
%%%%
\section{Replica trick proof of the linked-cluster (cumulant) expansion}  \label{app:replica}
The simplest proof of the linked-cluster theorem follows from the replica trick.\cite{Negele} The starting point is the expression for $W(t)$ from Eq.~(\ref{eq:W_contour}). We then introduce a set of $R$ replicas of the nuclear system (i.e.~$R$ independent quantum dots). Since the generalized (contour ordered)  exponent still has the defining property that $e^{X+Y} \! =\! e^{X}e^{Y}$ for independent (commuting) operators $X$ and $Y$, the $R$-th power of the original $W$ is simply equal to the $W(t)$ calculated in the space of all the replicas. 
The Hamiltonian in this space is then simply the sum of the Hamiltonians corresponding to $R$ uncoupled systems, and the interaction in Eq.~(\ref{eq:W_contour}) becomes $\mathcal{\hat{V}} \! = \! \sum_{r,r'}^{R} \delta_{r,r'} \mathcal{\hat{V}}_{r}$, i.e.~it is diagonal in replica indices $r$, $r'$. Now we use the fact that
\beq
W^{R} = e^{R \ln W} = 1 + R \ln W + \sum_{m=2}^{\infty} \frac{(R\ln W)^{m}}{m!} \,\, ,
\eeq
so that the part of $W^{R}$ which is linearly proportional to $R$ is simply $\ln W$. When we look at the ring diagrams derived using the replicated interaction, we see that since the different replicas are independent, each linked (ring) diagram has to be summed over an additional label $r$. Due to the equivalence of different copies of the system, each ring diagram acquires a factor of $R$. Consequently, the diagrams which are products of $k$ rings acquire a prefactor of $R^{k}$. Therefore, $\ln W$ is given by a sum of all the linked terms in the perturbation expansion of $W$.

%\bibliography{../../refs_quant,../../refs_doktorat}

\begin{thebibliography}{88}
\bibitem[{foo()}]{footnote_secular}
\bibinfo{note}{The spectral diffusion calculations using the cluster technique
  from Ref.~\onlinecite{Witzel_PRB06} can be as well performed using the full
  dipolar interaction with nonsecular terms included, and for magnetic fields
  under consideration here the results are practically unchanged.}
\end{thebibliography}

\end{document}